%% file: lmcs2023.tex
\documentclass{lmcs}
\pdfoutput=1
\usepackage[utf8]{inputenc}

\usepackage{lastpage}
\lmcsdoi{21}{4}{15}
\lmcsheading{}{\pageref{LastPage}}{}{}%
{May~14,~2024}{Oct.~30,~2025}{}


\usepackage[T1]{fontenc}
\usepackage[main=english]{babel} %typography check
\usepackage{amssymb, amsmath, amsfonts, graphicx} %math packages
\usepackage{float}
\usepackage{hyperref} %for hyperlinks
\usepackage[makeroom]{cancel} %for crossing out letters
\usepackage{xspace} %extra space after a macro
\usepackage{comment} %to comment the code
\usepackage{float} %to insert figures where I want
\usepackage{array} %arrays and tables 
\usepackage{subcaption} %subfigures
\usepackage{enumitem}  %new features to do with lists
\usepackage{pdfpages} %simplifies to insert an external pdf document
\usepackage{lipsum} %lorem ipsum
\usepackage[noadjust]{cite} %citation design
\usepackage{afterpage} %in order not to overload the float mechanism
\usepackage{tabularx} %to manipulate the table width
\usepackage{cleveref} %clever references
\usepackage{caption} %in order to manipulate the size of the caption font
\usepackage{paralist}
\usepackage{microtype}
\usepackage{xcolor}

\input{mymacros}

\def\eg{{\em e.g.}}

\begin{document}

\title[On guarded extensions of MMSNP]{On guarded extensions of MMSNP}
\titlecomment{An extended abstract of this paper has appeared at CiE 2023.}
\thanks{This work was supported by ANR project DIFFERENCE (\url{https://anr.fr/Projet-ANR-20-CE48-0002}) and by the European Union (ERC, POCOCOP, 101071674). Views and opinions expressed are however those of the author(s) only and do not necessarily reflect those of the European Union or the European Research Council Executive Agency. Neither the European Union nor the granting authority can be held responsible for them.}  %optional

% affiliations are numbered automatically with a, b, c (see below)
% use the optional argument to indicate the affiliation(s) of each author
% omit the argument if there is only one author, or only one affiliation
\author[A.~Barsukov]{Alexey Barsukov\lmcsorcid{0000-0001-7627-4823}}[a]
\author[F.~Madelaine]{Florent R. Madelaine\lmcsorcid{0000-0002-8528-7105}}[b]

% affiliation 1 (automatically numbered a)
\address{Department of Algebra, Faculty of Mathematics and Physics, Charles University, Sokolovsk{\'a} 49/83, 186 00 Prague 8, Czechia}  %optional
% write emails for all authors having that affiliation
\email{alexey.barsukov@matfyz.cuni.cz}  %optional

% affiliation 2 (automatically numbered b)
\address{Univ Paris Est Creteil, LACL, F-94010 Creteil, France}  %optional
\email{florent.madelaine@u-pec.fr}  %optional

\begin{abstract}
Feder and Vardi showed that the class \emph{Monotone Monadic SNP without inequality} (MMSNP) has a P vs NP-complete dichotomy if and only if such a dichotomy holds for finite-domain \emph{Constraint Satisfaction Problems} (CSPs).
Moreover, they showed that none of the three classes obtained by removing one of the defining properties of MMSNP (monotonicity, monadicity, no inequality) has a dichotomy.
The overall objective of this paper is to study the gaps between MMSNP and each of these three superclasses, where the existence of a dichotomy remains unknown.
For the gap between MMSNP and Monotone SNP without inequality, we study the class \emph{Guarded Monotone SNP without inequality} (GMSNP) introduced by Bienvenu, ten Cate, Lutz, and Wolter, and prove that GMSNP has a dichotomy if and only if a dichotomy holds for GMSNP problems over signatures consisting of a unique relation symbol.
For the gap between MMSNP and MMSNP with inequality, we introduce a new class \emph{MMSNP with guarded inequality}, that lies between MMSNP and MMSNP with inequality and that is strictly more expressive than the former and still has a dichotomy.
For the gap between MMSNP and Monadic SNP without inequality, we introduce a logic that extends the class of Matrix Partitions in a similar way how MMSNP extends finite-domain CSP, and pose an open question about the existence of a dichotomy for this class.
Finally, we revisit the theorem of Feder and Vardi, which claims that the class NP embeds into MMSNP with inequality.
We give a detailed proof of this theorem as it ensures no dichotomy for the right-hand side class of each of the three gaps.
\end{abstract}

\maketitle

\section{Introduction}

%%%%%%%%%%%%%%%%%%%%%%%%%%%%%%%%%%%%%%%%%%%%%%%%%%%%%%%%%%%%%%%%%%%%%%%%%%%%%%
% A paragraph about the history of the dichotomy question
%%%%%%%%%%%%%%%%%%%%%%%%%%%%%%%%%%%%%%%%%%%%%%%%%%%%%%%%%%%%%%%%%%%%%%%%%%%%%%

\subsection{The dichotomy question}

A subclass of NP has a \emph{dichotomy} if every problem in this class is either solvable in polynomial time or NP-complete.
A consequence of Ladner's theorem~\cite{ladner1975} is that if P is different from NP, then NP does not have a dichotomy.
This gave motivation to study subclasses of NP with regard to the existence of a dichotomy.
A large class of problems, which was believed to have this property was the class of finite-domain \emph{Constraint Satisfaction Problems} (CSP).
The existence of a dichotomy for finite-domain CSP was conjectured by Feder and Vardi~\cite{federvardi1998}.
Firstly, this conjecture was supported by several results confirming it for subclasses such as CSPs on a two-element domain~\cite{schaefer1978} and CSPs on undirected graphs~\cite{hellnesetril1990}.
Secondly, it generalised the conjecture of Bang-Jensen and Hell that a dichotomy holds for CSPs on directed graphs, where the template is a \emph{smooth} digraph, i.e., the in- and out- degree of every vertex is greater than zero~\cite{bangjensen_hell1990}.
Today, both conjectures are known to be true: the weaker version for smooth digraphs was proved by Barto, Kozik, and Niven~\cite{barto_kozik_niven2009}, and the stronger version was proved independently by Bulatov~\cite{bulatov2017} and Zhuk~\cite{zhuk2020}.

Apart from conjecturing a dichotomy for CSPs, Feder and Vardi~\cite{federvardi1998} built a link between NP and CSP through descriptive complexity.
By Fagin's theorem~\cite{fagin1974}, the class NP is captured by \emph{Existential Second Order logic} (ESO) which means that a class of finite structures is definable by a sentence in ESO if and only if it is decidable by a nondeterministic Turing machine running in polynomial time.
Feder and Vardi studied subclasses of the logic \emph{SNP} that itself is a fragment of ESO.
They considered three syntactic properties that restrict the class SNP:
\begin{compactitem}
\item \emph{monotone} -- all input relations have odd nested number of negations,
\item \emph{monadic} -- all existentially quantified relations have arity at most one,
\item \emph{without inequality} -- the ``='' relation and its negation are not allowed.
\end{compactitem}
Their main result states that, under these three restrictions, the resulting fragment of SNP, called \emph{Monotone Monadic SNP without inequality} (MMSNP), is contained in CSP modulo polynomial-time equivalence.
Moreover, they showed that it is necessary to have all the three properties in order to have a dichotomy, because removing any of these properties would give a class that does not have a dichotomy.

%%%%%%%%%%%%%%%%%%%%%%%%%%%%%%%%%%%%%%%%%%%%%%%%%%%%%%%%%%%%%%%%%%%%%%%%%%%%%%
% A paragraph, where I argue that the current state of the art contains gaps
% between dichotomy and no dichotomy
%%%%%%%%%%%%%%%%%%%%%%%%%%%%%%%%%%%%%%%%%%%%%%%%%%%%%%%%%%%%%%%%%%%%%%%%%%%%%%

\subsection{Three gaps above MMSNP}
Feder and Vardi's result about MMSNP together with Bulatov and Zhuk's proofs of a dichotomy for CSP imply that the logic MMSNP also has a dichotomy.
As removing any of the three properties from MMSNP gives a class that has no dichotomy, there appear three gaps, in which the existence of a dichotomy is unknown.
Each gap consists of logical classes that contain MMSNP, where one of the three properties is relaxed.

\begin{figure}
\centering
\includegraphics[width=\textwidth]{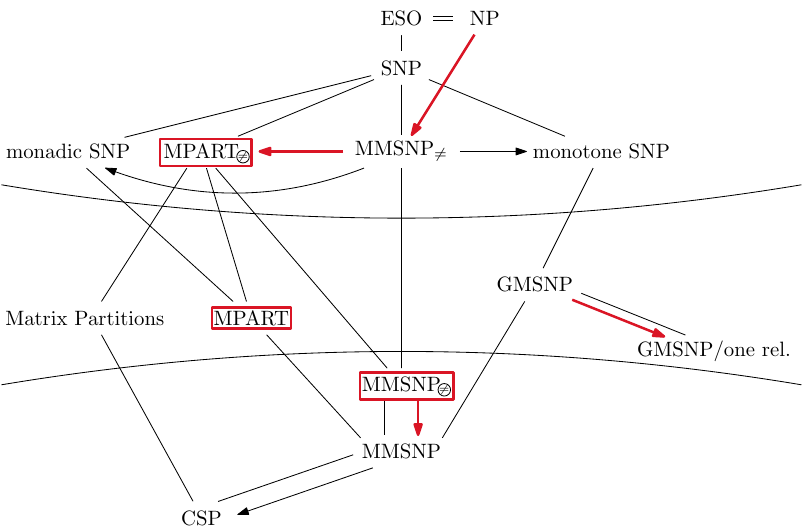}
\caption{Classes at the bottom exhibit a $\P$vs$\NPc$ dichotomy, while
  those at the  top do not, and it remains open for classes in the
  middle. Undirected edges stand for straightforward inclusions and directed edges stand for inclusions under polynomial-time reductions. The classes and inclusions discussed in this paper are highlighted with red. The sign $\guardedineq$  denotes classes extended with guarded inequality.}
\label{fig:diagram}
\end{figure}

The diagram in~\Cref{fig:diagram} displays the current state of the art.
There is an (undirected) edge between two classes if the class below is contained in the class above.
Directed edges stand for the containment modulo polynomial-time equivalence, i.e., for every problem of the class at the tail there exists a polynomial-time equivalent problem in the class at the head.
In \Cref{fig:diagram}, the class obtained by removing the ``monotone'' property is called \emph{monadic SNP}, removing ``monadicity'' gives the class \emph{monotone SNP}, while allowing ``inequality'' results in the class \emph{MMSNP$_{\not=}$}.

The contributions of the paper can be split in two parts.
In one part, we argue that none of the three upper bound classes has a dichotomy.
For this, we revisit Feder and Vardi's paper~\cite{federvardi1998} and give a detailed proof that every nondeterministic Turing machine is polynomial-time equivalent to a problem in monotone monadic SNP without inequality.
In the other part, we focus on the three gaps and study them from the dichotomy question perspective.
All the classes from the gaps are obtained from MMSNP by making one of its three properties less strict:
\begin{compactitem}
  \item ``monadicity'' is relaxed by allowing existentially quantified relations of any arity as long as they are guarded by input relational tuples;
  \item ``monotonicity'' is relaxed by allowing negated input relational atoms as long as any two input atoms in the same negated conjunct are either both negated or both non-negated;
  \item ``no inequality'' is relaxed by allowing $\not=$-atoms as long as they are guarded by input relational tuples.
\end{compactitem}
Each of these three relaxations is \emph{guarded}, in some sense: either by input relational atoms or by negated conjuncts.
This feature that all the classes in the gaps have in common, gave the title of this paper.

%%%%%%%%%%%%%%%%%%%%%%%%%%%%%%%%%%%%%%%%%%%%%%%%%%%%%%%%%%%%%%%%%%%%%%%%%%%%%%
% A paragraph about the "monadic relaxation" gap 
% (infinite CSPs, GMSNP)
%%%%%%%%%%%%%%%%%%%%%%%%%%%%%%%%%%%%%%%%%%%%%%%%%%%%%%%%%%%%%%%%%%%%%%%%%%%%%%

\subsubsection*{No dichotomy for MMSNP with inequality}

When Feder and Vardi proved that there is no dichotomy for the three logics extending MMSNP, they first showed an inclusion of NP into $\MMSNPineq$ modulo polynomial-time equivalence, and then they showed such inclusions of $\MMSNPineq$ into the other two classes.
In their paper, the proof of no dichotomy for $\MMSNPineq$ provides key ideas, in particular, the usage of \emph{oblivious Turing machines}, but it is very brief and thus not transparent enough.
In this paper we give a detailed proof of this statement, since it follows precisely from it that the right boundary of each interval has no dichotomy, which is a necessary criterion for studying the complexity of each of the three gaps.
Having an explicit proof of this statement makes the gaps \emph{well-defined}, i.e., the left-hand side class MMSNP has a dichotomy whereas all three right-hand side classes do not.
The interested reader may consult~\cite{barsukovphd} for extensive proofs regarding the two other logics: monotone SNP and monadic SNP.

In the following three paragraphs, we describe which classes each gap contains and what our contributions are.

\subsubsection*{Relaxing monadicity}

The gap between MMSNP and monotone SNP without inequality, among those three, is the most studied one.
The reason is that the family of finite models of every sentence $\Phi$ in monotone SNP without inequality is closed under taking inverse homomorphisms, i.e., if $\Phi$ holds in $\structure{B}$ and if $\structure{A}$ maps homomorphically to $\structure{B}$, then $\Phi$ also holds in $\structure{A}$.
For a family of finite structures over a finite relational signature, being closed under inverse homomorphisms and disjoint unions is equivalent to being the CSP of some (possibly infinite) structure (Lemma 1.1.8 in~\cite{bodirsky_book}).
CSPs with infinite domain templates contain the whole class NP modulo Turing equivalence (Corollary 13.2.3 in~\cite{bodirsky_book}).
Moreover, they can even be undecidable as, for example, solving Diophantine equations can be viewed as a CSP~\cite{matiyasevich}.

The largest fragment of infinite CSP, that is wholly contained in NP, consists of those CSPs, where the template is a (first-order) reduct of a finitely bounded homogeneous structure.
For this class, a dichotomy was conjectured by Bodirsky and Pinsker in 2011, see~\cite{bodirsky_pinsker_conjecture}.
The conjecture has been confirmed for many particular classes; for more information see the survey~\cite{pinsker_survey}.

One particularly interesting logical class that belongs to the gap between MMSNP and monotone SNP without inequality for which the dichotomy question is still open, is called \emph{Guarded Monotone SNP} (GMSNP).
It was introduced by Bienvenu, ten Cate, Lutz, and Wolter in~\cite{bienvenu2014}.
This class allows existentially quantified relations of arity greater than 1 but all such relations must be \emph{guarded} by input relations, i.e., the image of every existential-relational tuple must be contained in the image of some input-relational tuple.
In~\cite{bienvenu2014}, the authors showed, in particular, that GMSNP is strictly more expressive than MMSNP and that it is equally expressive as the class $\MMSNPtwo$ studied earlier by Madelaine~\cite{madelaine2009}.
The latter generalises MMSNP by allowing existentially quantified unary relations whose domain is both set of input vertices and relational tuples.
In~\cite{bodirsky_asnp}, it was shown that every problem in GMSNP is equivalent to a union (disjunction) of CSPs of first-order reducts of finitely bounded homogeneous structures.
This means that the resolution of Bodirsky-Pinsker conjecture would yield a dichotomy for GMSNP.
Conversely, showing a dichotomy for GMSNP would be a step towards proving the conjecture of Bodirsky and Pinsker.

In this paper, we show that it suffices to show a dichotomy only for those GMSNP problems, where the input relational signature consists of a single relation symbol.

%%%%%%%%%%%%%%%%%%%%%%%%%%%%%%%%%%%%%%%%%%%%%%%%%%%%%%%%%%%%%%%%%%%%%%%%%%%%%%
% A paragraph about the "monotone relaxation" gap
% (matrix partitions and logics that I introduce)
%%%%%%%%%%%%%%%%%%%%%%%%%%%%%%%%%%%%%%%%%%%%%%%%%%%%%%%%%%%%%%%%%%%%%%%%%%%%%%

\subsubsection*{Relaxing monotonicity}

There has already been some research done in generalizing CSPs in terms of modifying the notion of the homomorphism.
Typically, such problems are not closed under inverse homomorphisms.
One intensively studied such class is called \emph{Matrix Partitions}.
Every such problem is described by a square matrix $A$ of size $m\times m$ whose elements take values in the set $\{0,1,\ast\}$.
The input of the problem consists of graphs.
For a given input graph $\structure{G}$, the Matrix Partition problem parametrised by $A$ asks: is there a partition of the set of vertices of $\structure{G}$ in sets $P_1,\ldots,P_m$ such that for all $i,j\in\{1,\ldots,m\}$ and for every pair of \emph{distinct} vertices $v_i\in P_i,v_j\in P_j$,
\begin{compactitem}
\item $A_{ij}=0$ implies that there must be no edge between $v_i$ and $v_j$,
\item $A_{ij}=1$ implies that there must be an edge between $v_i$ and $v_j$,
\item $A_{ij}=\ast$ means that there is no restriction.
\end{compactitem}
In particular, if the $i$th diagonal element $A_{ii}$ is $0$, then $P_i$ induces an independent set, and if $A_{ii}=1$, then $P_i$ induces a clique.

Matrix Partitions contain all CSPs on graphs and the dichotomy question for them is open (Problem 2 in~\cite{hell2014}).
This containment is strict because Matrix Partitions, in general, are not closed under inverse homomorphisms.
For further reading, see the survey~\cite{hell2014}.
\begin{exa}
Consider the 3-colouring and the split graph recognition problems.
They can be seen as Matrix Partition problems for the matrices $A_3$ and $A_{\mathrm{split}}$, respectively, where
\begin{equation*}
A_3 = \begin{pmatrix}0 & \ast & \ast\\ \ast & 0 & \ast\\ \ast & \ast & 0\end{pmatrix}\qquad A_{\mathrm{split}} = \begin{pmatrix}0 & \ast\\ \ast & 1\end{pmatrix}.
\end{equation*}
\end{exa}

Another way to look at Matrix Partitions was given in~\cite{barsukovkante}.
There, it was proposed to see $k$-ary relations defined on a set $A$ not as subsets of $A^k$ but rather as mappings from $A^k$ to some fixed poset.
In this setting, a homomorphism between structures $\structure{A}$ and $\structure{B}$ was defined as a mapping between their domains $A$ and $B$ that sends every relational tuple to a tuple whose value is greater or equal (with respect to the partial order) than the value of the original tuple.
If the poset has elements $0,1$, and the order is defined by the inequality $0\preceq 1$, this is equivalent to usual CSPs.
If the poset has elements $0,1,\ast$, and the order is defined by $0\preceq \ast$ and $1\preceq \ast$, then this is \emph{almost} equivalent to Matrix Partitions.

These two definitions are not completely equivalent because, in Hell's definition, the partition constraints are applied to pairs of \emph{distinct} vertices whereas the class studied in~\cite{barsukovkante} does it for all pairs.
As a result, trying to write down a problem from the former class as an SNP sentence would require the use of inequalities as input relations while the latter class does not have this issue.
It remains unclear whether these two classes are polynomial-time equivalent.
However, we feel that the variation from~\cite{barsukovkante} is more amenable to the framework of descriptive complexity while retaining the key property of exhibiting non-monotonicity in a suitably restricted form that could prevent NP-intermediate complexity to arise.

In the present article, we introduce a logic MPART that lies in the gap between MMSNP and monadic SNP without inequality.
The class MPART extends MMSNP in a similar fashion as how Matrix Partitions extend graph colourings.
However, using the inequality as an input relation symbol is not allowed (which is required to belong to the gap).
Moreover, we show that, adding inequality to $\MPART$ in order to express Hell's Matrix Partitions, results in a logic, denoted by $\GMPARTineq$, that contains $\MMSNPineq$ modulo polynomial-time equivalence and, therefore, has no dichotomy.

%%%%%%%%%%%%%%%%%%%%%%%%%%%%%%%%%%%%%%%%%%%%%%%%%%%%%%%%%%%%%%%%%%%%%%%%%%%%%%
% A paragraph about the "inequality relaxation" gap
% (matrix partitions and logics that I introduce)
%%%%%%%%%%%%%%%%%%%%%%%%%%%%%%%%%%%%%%%%%%%%%%%%%%%%%%%%%%%%%%%%%%%%%%%%%%%%%%

\subsubsection*{Relaxing ``no inequality''}

To our knowledge, there are no classes from this gap that have been studied in the literature.
We introduce a new logical class called \emph{MMSNP with guarded inequality} ($\GMMSNPineq$) that belongs to the gap.
It generalises MMSNP by adding a restricted form of inequality, where each $\not=$-atomic formula is guarded by an atomic formula of some input relational symbol, similarly as atomic formulas of existential relations are guarded by input relational atomic formulas in GMSNP.
We show that this class is strictly more expressive than MMSNP.
However, it is still included in MMSNP modulo polynomial-time equivalence and, therefore, it has a dichotomy.

%%%%%%%%%%%%%%%%%%%%%%%%%%%%%%%%%%%%%%%%%%%%%%%%%%%%%%%%%%%%%%%%%%%%%%%%%%%%%%
% Outline
%%%%%%%%%%%%%%%%%%%%%%%%%%%%%%%%%%%%%%%%%%%%%%%%%%%%%%%%%%%%%%%%%%%%%%%%%%%%%%

\subsection{Outline}

We give all the necessary definitions in Section~\ref{section:preliminaries}.
Then, in Section~\ref{section:gaps}, we visit the three gaps one by one.
In Subsection~\ref{section:gmmsnpineq}, we introduce $\MMSNP$ with guarded inequality ($\GMMSNPineq$) into the gap between MMSNP and MMSNP with inequality and prove that it exhibits a dichotomy.
In Subsection~\ref{section:gmsnp}, we study the class $\GMSNP$ that is in the gap between $\MMSNP$ and monotone SNP without inequality.
We establish that proving a dichotomy for signatures consisting of a single input relation would suffice.
In Subsection~\ref{section:mp}, we give the definition for the logics $\MPART$ and $\GMPARTineq$ and show that the latter does not have a dichotomy.
Finally, in Section~\ref{section:snp_subclasses}, we provide a complete proof of one crucial step of Feder and Vardi's argument to establish that the three stronger siblings of $\MMSNP$ cannot have a dichotomy unless $\P$ equals $\NP$, which makes the three gaps well-defined.

\section{Preliminaries}\label{section:preliminaries}

Denote by $\tau$ a finite relational signature.
Let $\sigma$ be a set of relation symbols $\{ \rel{X}_1,\ldots,  \rel{X}_s\}$.
Relation symbols of $\tau$ are called \emph{input} and relation symbols of $\sigma$ are called \emph{existential}. 

Every $\tau$-sentence $\Phi$ in $\SNP$ can be written in the following form:
\begin{equation}\label{eq:negated_cnf}
\exists \rel{X}_1,\ldots,  \rel{X}_s\;\forall\tuple{x} \; \bigwedge_i \neg (\alpha_i\wedge\beta_i\wedge \epsilon_i),
\end{equation}
where $\alpha_i$ is a conjunction of atomic or negated atomic $\tau$-formulas, $\beta_i$ is a conjunction of atomic or negated atomic $\sigma$-formulas, and $\epsilon_i$ is a conjunction of inequalities $x_i\not=x_j$ such that $x_i,x_j\in\tuple{x}$.
In this paper, an $\SNP$ sentence written in the form of \cref{eq:negated_cnf} is said to be in \emph{standard} form.

While SNP sentences are denoted by upper-case letters: $\Phi$, their quantifier-free parts are denoted by lower-case letters: $\phi(\tuple{x})$.
Tuples of first-order variables or elements are denoted with bold lower-case letters: $\tuple{x},\tuple{a},\dots$ 

For the sake of brevity, we write ``atom'' instead of ``atomic formula''.
Every subformula $\neg (\alpha_i\wedge\beta_i\wedge\epsilon_i)$ of the quantifier-free part $\phi(\tuple{x})$ is called a \emph{negated conjunct}.

A $(\tau\cup\sigma)$-structure $\structure{A}^\sigma$ is a \emph{$\sigma$-expansion} of a $\tau$-structure $\structure{A}$ if these two structures have the same domain and if, for every $\rel{R}$ in $\tau$, the relations $\rel{R}^{\structure{A}}$ and $\rel{R}^{\structure{A}^\sigma}$ are the same.
If $\structure{A}^\sigma$ is a $\sigma$-expansion of $\structure{A}$, then $\structure{A}$ is called the \emph{$\tau$-reduct} of $\structure{A}^\sigma$.
For a $\tau$-sentence $\Phi$ in standard form, we say that $\Phi$ \emph{holds} in a $\tau$-structure $\structure{A}$, denoted $\structure{A}\models\Phi$, if there exists a $\sigma$-expansion $\structure{A}^\sigma$ such that $\structure{A}^\sigma\models\phi(\tuple{a})$, for every assignment $\tuple{a}$ of elements of $\structure{A}$ to the variables of $\phi(\tuple{x})$.
When we want to highlight that a tuple $\tuple{a}$ of elements of a set $A$ is assigned to the tuple $\tuple{x}$ of variables, we write $\tuple{a}\in A^{\tuple{x}}$.

In this paper, both the set of all finite $\tau$-structures, where $\Phi$ holds, and the problem deciding the satisfiability of $\Phi$, are denoted by $\Sat(\Phi)$.

An $\SNP$-sentence $\Phi$ is called \emph{monadic} if all existential symbols have arity one.
It is called \emph{monotone} if every $\alpha_i$ omits negated $\tau$-atoms, and \emph{without inequality} if every $\epsilon_i$ is empty.

Alternatively, one can think of being ``monotone'' and ``without inequality'' as of being closed under
inverse homomorphisms. The latter means that a structure $\structure{A}$ satisfies $\Phi$ whenever $\structure{A}$ homomorphically maps to a structure $\structure{B}\in\Sat(\Phi)$.
Indeed, every sentence in monotone SNP without inequality is closed under inverse homomorphisms and the other direction is provided by the following.
\begin{thmC}[\cite{federV03}]\label{th:monotone-hom-closed} 
For every sentence $\Phi$ in $\SNP$, the class $\Sat(\Phi)$ is closed under inverse homomorphisms only if $\Phi$ is logically equivalent to a sentence in monotone SNP without inequality.
\end{thmC}

Denote by $\MMSNP$ the fragment of $\SNP$ consisting of sentences that are monotone, monadic, and without inequality.

An $\SNP$ sentence $\Phi$ in standard form is in \emph{Guarded Monotone Strict NP} ($\GMSNP$) if
\begin{inparaenum}[(a)]
\item each $\alpha_i$ is a conjunction of non-negated $\tau$-atoms,
\item each $\beta_i$ is a conjunction of $\sigma$-atoms or negated $\sigma$-atoms, and
\item each $\epsilon_i$ is empty.
\end{inparaenum}
Also, for every negated $\sigma$-atom $\neg \rel{X}(\tuple{x}_1)$ from every $\beta_i$ there exists either
\begin{inparaenum}[(i)]
\item a $\tau$-atom $\rel{R}(\tuple{x}_2)$ from $\alpha_i$, or
\item a non-negated $\sigma$-atom $\rel{X}'(\tuple{x}_2)$ from $\beta_i$
\end{inparaenum}  
such that $\tuple{x}_1\subseteq\tuple{x}_2$.
In this case, we say that $\rel{R}(\tuple{x}_2)$ (or $\rel{X}'(\tuple{x}_2)$) \emph{guards} $\rel{X}(\tuple{x}_1)$.

\begin{defi}\label{def:gmmsnpineq} 
An $\SNP$ sentence $\Phi$ in standard form is in \emph{$\MMSNP$ with guarded inequality} ($\GMMSNPineq$) if
\begin{inparaenum}[(a)]
\item each $\alpha_i$ is a conjunction of non-negated $\tau$-atoms,
\item each $\beta_i$ is a conjunction of unary $\sigma$-atoms or negated unary $\sigma$-atoms, and
\item for every inequality $x_j\not=x_j'$ of each $\epsilon_i$ there exists a $\tau$-atom in $\alpha_i$ that contains both $x_j$ and $x_j'$.
\end{inparaenum}
\end{defi}

\begin{rem}\label{remark:yes_no_instance}
For every sentence that we consider, we assume that it is not trivial, \ie, it has some YES and some NO instances.
\end{rem}

%%%%%%%%%%%%%%%%%%%%%%%%%%%%%%%%%%%%%%%%%%%%%%%%%%%%%%%%%%%%%%%%%%%%%%%%%%%%%%%%
\section{Studying the three gaps}\label{section:gaps}
%%%%%%%%%%%%%%%%%%%%%%%%%%%%%%%%%%%%%%%%%%%%%%%%%%%%%%%%%%%%%%%%%%%%%%%%%%%%%%%%
\subsection{MMSNP with guarded inequality}\label{section:gmmsnpineq}

In this subsection, we prove the following theorem.
\begin{thm}\label{th:gmmsnpineq_embeds_mmsnp} 
For every sentence $\Phi$ in $\GMMSNPineq$ there exists a sentence $\Phi'$ in $\MMSNP$ such that the problems $\Sat(\Phi)$ and $\Sat(\Phi')$ are polynomial-time equivalent.
\end{thm}

For the sake of brevity, we consider the case when $\tau = \{\rel{R}\}$, where $\rel{R}$ has
arity $n$. 
In the case of several relations, one can apply the same procedure for each relation independently.

The construction has two main steps.
At first, we enrich every negated conjunct of $\Phi$ with inequalities while keeping the resulting sentence logically equivalent to the initial one.
This transformation is the same as the one used in Lemma 4 in \cite{federV03}.
After that, we construct an equivalent $\MMSNP$ sentence $\Phi'$ by changing the input relational signature.
Each new relation symbol is associated with an equivalence relation on an $n$-element set, where $n$ is the arity of $\rel{R}$.

\subsection*{Enriching the sentence with inequalities} 
For every negated conjunct $\neg\phi$ of  $\Phi$ and for every two distinct variables $x,y$ that appear in the same $\rel{R}$-atom of $\neg\phi$,  replace this negated conjunct with two negated conjuncts: the first one is obtained from $\neg\phi$ by adding the inequality $x\not=y$ to the conjunction, the second one is obtained from $\neg\phi$ by replacing every occurrence of the variable $y$ with the variable $x$.
Repeat this procedure until every negated conjunct contains the inequality $x\not=y$ for every two distinct variables $x,y$ of this conjunct. Call the resulting sentence $\Psi$.

\begin{exa} \label{example:enriching_inequalities}
If $\Phi$ contains a negated conjunct $\neg\bigl(\rel{R}(x,y,z)\wedge\rel{X}_1(x)\wedge\rel{X}_2(y)\bigr)$, then it is replaced by the following five negated conjuncts: 
\begin{align*}
&\neg\bigl(\rel{R}(x,y,z)\wedge\rel{X}_1(x)\wedge\rel{X}_2(y)\wedge x\not=y \wedge x\not=z\wedge y\not=z\bigr)\\
\wedge &\neg\bigl(\rel{R}(x,x,y)\wedge\rel{X}_1(x)\wedge\rel{X}_2(x)\wedge x\not=y\bigr)\wedge \neg\bigl(\rel{R}(x,y,x)\wedge\rel{X}_1(x)\wedge\rel{X}_2(y)\wedge x\not=y\bigr)\\
\wedge &\neg\bigl(\rel{R}(x,y,y)\wedge\rel{X}_1(x)\wedge\rel{X}_2(y)\wedge x\not=y\bigr)\wedge \neg\bigl(\rel{R}(x,x,x)\wedge\rel{X}_1(x)\wedge\rel{X}_2(x)\bigr).
\end{align*}
\end{exa}

The following lemma is a direct consequence of the construction.
\begin{lem}\label{lemma:gmmsnpineq_enrich_ineq} 
Every $\GMMSNPineq$ sentence $\Phi$ is logically equivalent to a $\GMMSNPineq$ sentence $\Psi$ such that, for every negated conjunct $\neg\psi_i$ of $\Psi$ and for every two different variables $x,y$ that appear within some $\rel{R}$-atom of $\psi_i$, this conjunct contains the inequality $x\not=y$.
\end{lem}
\proof
Let $\Phi$ and $\Psi$ be as in the paragraph above Example~\ref{example:enriching_inequalities}, and let $\structure{A}$ be a $\tau$-structure.
Suppose that $\structure{A}\not\models\Phi$; then, for every $\sigma$-expansion $\structure{A}^\sigma$ there exists a negated conjunct $\neg\phi_i(\tuple{x})$ of $\Phi$ and a tuple $\tuple{a}\in A^\tuple{x}$ such that $\structure{A}^\sigma\models\phi_i(\tuple{a})$.
Let $\neg\psi_i(\tuple{x}')$ be the negated conjunct of $\Psi$ that is obtained from $\neg\phi_i(\tuple{x})$ by adding the inequality $x_i\not=x_j$, for every two distinct elements $a_i,a_j\in\tuple{a}$, and by identifying $x_i$ with $x_j$, for every two elements $a_i,a_j\in\tuple{a}$ that are the same.
By the construction of $\Psi$, such a conjunct exists, so it will not be satisfied in $\structure{A}^\sigma$.
If $\structure{A}\not\models\Psi$, then, for every $\sigma$-expansion $\structure{A}^\sigma$, there exists a negated conjunct $\neg\psi_i$ of $\Psi$ that is false for some assignment.
It is associated with some negated conjunct $\neg\phi_i$ of $\Phi$, and, by construction, $\phi_i$ will not be satisfied in $\structure{A}^\sigma$ for the same assignment.
We conclude that $\Phi$ and $\Psi$ are logically equivalent.
\qed

\medskip

\begin{figure}[ht]
\centering
\includegraphics[width=\textwidth]{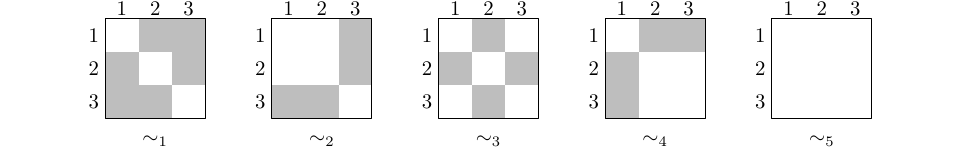}
\caption{All the 5 equivalence relations on a $3$-element set.}
\label{fig:example_ternary}
\end{figure}

Now, by Lemma~\ref{lemma:gmmsnpineq_enrich_ineq}, we may assume that, for every two variables $x,y$ that appear within the same $\rel{R}$-atom in some negated conjunct of $\Phi$, this negated conjunct contains $x\not=y$. 
For $n$ in $\mathbb{N}$, denote by $B_n$ the number of equivalence relations on a set of $n$ elements: $\sim_1,\ldots,\sim_{B_n}$. 
For each $\sim_k$, denote by $n_k:=\bigl|\{1,\ldots,n\}/\sim_k\bigr|$ the number of equivalence classes of this relation.
Every $n$-tuple $(x_1,\ldots,x_n)$ is associated with exactly one $\sim_k$ such that $x_i=x_j$ if and only if $i\sim_k j$.
If $\sim_k$ is associated with $\tuple{x}$, then we say that $\tuple{x}$ has \emph{equivalence type $k$}.
For every equivalence class $\{x_{c_1},\ldots, x_{c_l}\}$ of $\sim_k$, denote it by $[c]_{\sim_k}$, where $c = \min\{c_1,\ldots,c_l\}$ -- the smallest number of this set.
Then, introduce a linear ordering $\prec_k$ on the set $\{1,\ldots,n\}/\sim_k$ by setting $[i]_{\sim_k}\prec_k [j]_{\sim_k}$ if $i<j$. 
\begin{defi}\label{def:p} 
For every set $X$, define a function $\func{p}\colon \biguplus_{n=1}^\infty X^n \to \biguplus_{n=1}^\infty X^n$ as follows.
Let $\tuple{x}=(x_1,\ldots,x_n)$ in $ X^n$ be an $n$-tuple of equivalence type $k$.
Let $[s_1]_{\sim_k},\ldots,[s_{n_k}]_{\sim_k}$ be the $\sim_k$-equivalence classes such that $[s_i]_{\sim_k}\prec_k[s_j]_{\sim_k}$ if and only if $i<j$.
Then, put $\func{p}(\tuple{x}):= \bigl(x_{s_1},\ldots,x_{s_{n_k}}\bigr)$. 
That is, the function $\func{p}$ removes an element from a tuple if it is not its first occurrence.
\end{defi}
\begin{exa}
Consider a $3$-tuple $\tuple{t} = (y,x,x)$, it is associated with the equivalence relation $\sim_4$ from \Cref{fig:example_ternary}, the equivalence classes of $\sim_4$ on the set $\{1,2,3\}$ are $[1]_{\sim_4}=\{1\}$ and $[2]_{\sim_4} = \{2,3\}$.
Then, $\func{p}(\tuple{t})=(y,x)$ because $y$ is on the first coordinate of $\tuple{t}$, $x$ is on the second, and $[1]_{\sim_4}\prec_4[2]_{\sim_4}$.
\end{exa}

\subsection*{Constructing the $\MMSNP$ sentence} 
The $\MMSNP$ sentence $\Phi'$ has the same existential signature $\sigma$ as $\Phi$, however the input signature $\tau'$ is not the same.
Signature $\tau'$ has $B_n$-many relation symbols $\rel{R}_1,\ldots,\rel{R}_{B_n}$.
Each $\rel{R}_i$ is associated with one of the $B_n$ equivalence relations $\sim_k$ on the set $\{1,\ldots,n\}$, the arity of $\rel{R}_k$ equals to the number $n_k$ of equivalence classes of $\sim_k$. For example, if $n=3$, then $\tau'$ contains $B_3=5$ relation symbols $\rel{R}_1(\cdot,\cdot,\cdot), \rel{R}_2(\cdot,\cdot), \rel{R}_3(\cdot,\cdot), \rel{R}_4(\cdot,\cdot), \rel{R}_5(\cdot)$, see Figure~\ref{fig:example_ternary}.
We now describe how to construct the $\MMSNP$ sentence $\Phi'$ from $\Phi$.
\begin{compactitem} 
\item Firstly, remove all inequalities from $\Phi$. 
\item Then, replace every $\tau$-atom $\rel{R}(\tuple{x})$ with $\rel{R}_k\bigl(\func{p}(\tuple{x})\bigr)$, where $k$ is the equivalence type of $\tuple{x}$.
\item Finally, in order to simulate the inequalities that we have removed, we forbid the same variable to appear within $\tau'$-atoms more than once. For every $\rel{R}_k$ in $\tau'$ and for every $i<j$ in $[n_k]$, add to $\Phi'$ the following: 
\begin{equation}\label{eq:negated_conjunct_rigid} 
\neg\rel{R}_k(x_1,\ldots,x_{i-1},x,x_{i+1},\ldots,x_{j-1},x,x_{j+1},\ldots,x_{n_k}). 
\end{equation}
\end{compactitem}
\proof[Proof of Theorem~\ref{th:gmmsnpineq_embeds_mmsnp}] 
Let $\Phi'$ be obtained from $\Phi$ as above.
Let $\structure{A}$ be a $\tau$-structure.
Firstly, we will construct a $\tau'$-structure $\structure{A}'$ such that $\structure{A}\models\Phi$ if and only if $\structure{A}'\models\Phi'$.
The structure $\structure{A}'$ has the same domain $A'=A$.
For every $\rel{R}_k$ in $\tau'$ and for every tuple $\tuple{a} = (a_1,\ldots,a_n)$ of equivalence type $k$:
\begin{equation}\label{eq:construction_of_adash}
\rel{R}_k^{\structure{A}'}\bigl(\func{p}(\tuple{a})\bigr) \longleftrightarrow \rel{R}^\structure{A}(\tuple{a}).
\end{equation}
For every $\tau'$-structure there exists a $\tau$-structure $\structure{A}$ satisfying (\ref{eq:construction_of_adash}) if and only if all conditions of the type from (\ref{eq:negated_conjunct_rigid}) hold.
Otherwise, such a structure does not satisfy $\Phi'$, and we can output a fixed NO instance of $\Phi$ which always exists by Remark~\ref{remark:yes_no_instance}.

By construction, $\structure{A}'$ satisfies every negated conjunct from (\ref{eq:negated_conjunct_rigid}).
Let $\structure{A}^\sigma$ and $\structure{A}'^\sigma$ be $\sigma$-expansions such that, for every $\rel{X}$ in $\sigma$ and $a$ in $ A$,  $\rel{X}^{\structure{A}^\sigma}=\rel{X}^{\structure{A}'^\sigma}$.

Suppose that $\structure{A}^\sigma$ does not satisfy the first-order part of $\Phi$; then there exist a negated conjunct $\neg\phi_i(\tuple{x})$ of $\Phi$ and a tuple $\tuple{a}$ of $A^\tuple{x}$ such that $\structure{A}^\sigma\models\phi_i(\tuple{a})$. 
For every $\sigma$-atom $\rel{X}(a)$ of $\phi_i(\tuple{a})$, we have $\structure{A}^\sigma\models\rel{X}(a)$, which implies that $\structure{A}'^\sigma\models\rel{X}(a)$. 
For every $\tau$-atom $\rel{R}(\tuple{b})$ of $\phi_i(\tuple{a})$, we have $\structure{A}^\sigma\models\rel{R}(\tuple{b})$, which implies that $\structure{A}'^\sigma\models\rel{R}_k\bigl(\func{p}(\tuple{b})\bigr)$, by (\ref{eq:construction_of_adash}).
Thus, $\structure{A}'^\sigma\models\phi_i'(\tuple{a})$.
The other direction is similar.
\qed
\medskip

In the following, we show that the class $\GMMSNPineq$ strictly contains $\MMSNP$.
\begin{prop}
The class of $\tau$-sentences in $\GMMSNPineq$ is strictly more expressive than the class of $\tau$-sentences in $\MMSNP$.
\end{prop}
\proof
Let $\tau=\{\rel{E}(\cdot,\cdot)\}$ and let $\Phi$ be the sentence $\forall x,y\; \neg\bigl(\rel{E}(x,y)\wedge x\not=y\bigr)$.
This sentence is in $\GMMSNPineq$, however it is not logically equivalent to any $\MMSNP$ sentence because $\Sat(\Phi)$ is not closed under inverse homomorphisms.
\qed

\begin{rem}
While adding guarded inequalities makes $\MMSNP$ strictly more expressive, it preserves $\GMSNP$.
Let $\Phi$ be a $\tau$-sentence in $\GMSNP$ with guarded inequalities.
Firstly, introduce a new existential binary relation $\rel{Ineq}$.
Then, for every $\rel{R}\in\tau$ of arity $k$ and for every $i\in[k]$, add to $\Phi$ the negated conjunct $\neg\bigl(\rel{Ineq}(x_i,x_i)\wedge \rel{R}(x_1,\ldots,x_k)\bigr)$.
Finally, replace every inequality $x\not=y$ with $\rel{Ineq}(x,y)$.
The resulting sentence will be both in $\GMSNP$ and logically equivalent to $\Phi$.
\end{rem}

%%%%%%%%%%%%%%%%%%%%%%%%%%%%%%%%%%%%%%%%%%%%%%%%%%%%%%%%%%%%%%%%%%%%%%%%%%%%%%%%
\subsection{Guarded Monotone SNP over one-element signatures}\label{section:gmsnp}

In this section, we prove the following theorem.

\begin{thm}\label{th:gmsnp} 
For every finite relational signature $\tau$ there exists a relational signature $\tau_1$ that consists of a single relation symbol such that the class of $\GMSNP$ $\tau$-sentences has a dichotomy if and only if the class of $\tau_1$-sentences in $\GMSNP$ has a dichotomy.
\end{thm}

The single relation symbol is obtained by concatenating the original relation symbols, so its arity will be the sum of the arities of the original relations.
The new relation $\rel{P}$ will hold on a tuple $(\tuple{a}_1,\ldots,\tuple{a}_t)$ if and only if $\rel{R}_i$ holds on $\tuple{a}_i$ for all $i\in[t]$.
\subsubsection*{Why the obvious strategy fails}
It is not possible to apply this transformation right away for two reasons:
\begin{compactenum}[(a)]
\item There exists $\rel{R}\in\tau$ such that the original sentence rejects all input structures that contain $\rel{R}$-tuples, that is, the sentence contains the conjunct $\neg\rel{R}(\tuple{x})$.
Applying the transformation to an input structure, where $\rel{R}$ is empty, returns a structure, where $\rel{P}$ is empty.
So, the resulting input structure will be trivial while the original is not.
\item There exists $\rel{R}\in\tau$ and a forbidden conjunct of the original sentence $\Phi$ that does not contain $\rel{R}$-atoms. 
We cannot apply the transformation to such a conjunct because the resulting conjunct would not contain any $\rel{P}$-atoms, so it will become an empty constraint.
\end{compactenum}
For these two reasons, the proof consists of several stages.
In the first stage, we check if $\Phi$ holds in a structure consisting of a
single $\rel{R}$-tuple, for $\rel{R}$ in $\tau$. If not, then we remove this relation symbol because we need every relation of an input structure to be allowed to be non-empty.
In the second stage, we enrich every negated conjunct with $\tau$-atoms so that every relation symbol of $\tau$ is present in every conjunct.
In the last stage, we construct a suitable sentence over a signature containing a single relation symbol.

A relational structure is called \emph{connected} if it is not a disjoint union of two relational structures.
An $\SNP$ $\tau$-sentence $\Phi$ in the standard form is called \emph{connected} if, for every negated conjunct $\neg\phi(\tuple{x})$ of $\Phi$, the $(\tau\cup\sigma)$-structure $\structure{C}_\phi$ is connected, where the domain of $\structure{C}_\phi$ is the set of variables $\tuple{x}$ and, for every $\rel{R}\in\tau\cup\sigma$, $\rel{R}^{\structure{C}_\phi} = \{\tuple{y}\mid \rel{R}(\tuple{y})\;\text{is a non-negated atom of}\; \phi\}$.

The following lemma follows from the discussion after Proposition 1 in~\cite{bodirsky_asnp}.
\begin{lem}\label{lemma:gmsnp_connected} 
The following statements are equivalent.
\begin{compactenum}
\item For every $\tau$-sentence $\Phi$ in $\GMSNP$, $\Sat(\Phi)$ is either in P or NP-complete.
\item For every connected $\tau$-sentence $\Phi$ in $\GMSNP$, $\Sat(\Phi)$ is either in P or NP-complete. 
\end{compactenum}
\end{lem}
The following lemma is Proposition 1.4.11 in~\cite{bodirsky_book}.
\begin{lem}\label{lem:snp_disjoint_union} 
Let $\Phi$ be an $\SNP$ sentence.
Then, $\Sat(\Phi)$ is closed under disjoint unions if and only if $\Phi$ is logically equivalent to a connected $\SNP$ sentence.
\end{lem}
By Lemma~\ref{lemma:gmsnp_connected}, we can consider only connected $\tau$-sentences $\Phi$ in $\GMSNP$ and, by Lemma~\ref{lem:snp_disjoint_union}, $\Sat(\Phi)$ is closed under disjoint unions.

\subsubsection*{First stage: removing unnecessary relation symbols} 
For every $k$-ary $\rel{R}$ in $\tau$, let $\structure{B}_\rel{R}$ be the $\tau$-structure that consists of one $\rel{R}$-tuple $\tuple{b}_\rel{R} = (b_{\rel{R},1},\ldots,b_{\rel{R},k})$. That is, (a) $\structure{B}_\rel{R}$ has the domain $B_\rel{R}=\{b_{\rel{R},1},\ldots,b_{\rel{R},k}\}$, (b) $\rel{R}^{\structure{B}_\rel{R}}=\{\tuple{b}_\rel{R}\}$, and (c) for every other $\rel{S}\in\tau$, $\rel{S}^{\structure{B}_\rel{R}}=\varnothing$.
Denote by $\Phi^{\cancel{\rel{R}}}$ a connected $(\tau\setminus\{\rel{R}\})$-sentence that is obtained from $\Phi$ by removing all negated conjuncts that contain an $\rel{R}$-atom.

\begin{lem}\label{lemma:neg_r_case} 
If there exists $\rel{R}$ in $\tau$ such that a connected $\GMSNP$ $\tau$-sentence $\Phi$ does not satisfy $\structure{B}_\rel{R}$, then $\Sat(\Phi)$ is polynomial-time equivalent to $\Sat(\Phi^{\cancel{\rel{R}}})$.
\end{lem}
\proof
Suppose that $\structure{B}_\rel{R}\not\models\Phi$.
Let a $\tau$-structure $\structure{A}$ be an input instance of $\Sat(\Phi)$.
If $\rel{R}^\structure{A}\not=\varnothing$, then $\structure{B}_\rel{R}\to\structure{A}$, so, by Theorem~\ref{th:monotone-hom-closed}, $\structure{A}\not\models\Phi$, and we reduce it to some NO instance of $\Sat(\Phi^{\cancel{\rel{R}}})$.
Suppose that $\rel{R}^\structure{A}=\varnothing$; then we reduce $\structure{A}$ to its $(\tau\setminus\{\rel{R}\})$-reduct $\structure{A}^{\cancel{\rel{R}}}$. For every $\sigma$-expansion of $\structure{A}$, any negated conjunct with an $\rel{R}$-atom is always satisfied, as $\rel{R}^\structure{A}=\varnothing$.
Thus, $\structure{A}\models\Phi$ if and only if $\structure{A}^{\cancel{\rel{R}}}\models\Phi^{\cancel{\rel{R}}}$. For $\tau$-structures $\structure{A}$ such that $\rel{R}^\structure{A}=\varnothing$, the correspondence $\structure{A}\longleftrightarrow\structure{A}^{\cancel{\rel{R}}}$ is one-to-one.
This means that the two problems are polynomial-time equivalent.
\qed

A symbol $\rel{R}\in\tau$ is called \emph{redundant} for a connected $\GMSNP$ $\tau$-sentence $\Phi$ if $\Phi$ rejects all input structures that contain $\rel{R}$-tuples.
Repeatedly applying Lemma~\ref{lemma:neg_r_case}, we remove all redundant symbols and prove that, for every connected $\GMSNP$ sentence with  redundant symbols there exists a polynomial-time equivalent connected $\GMSNP$ sentence without redundant symbols.

\subsubsection*{Second stage: enriching negated conjuncts with $\tau$-atoms}\label{constr:non_empty}

Call a $\GMSNP$ sentence $\Phi$ \emph{enriched} if, for every negated conjunct $\neg\phi_i$ of $\Phi$ and every $\rel{R}\in\tau$, $\neg\phi_i$ contains at least one positive $\rel{R}$-atom.
We now show, for every connected $\Phi$ without redundant symbols, how to obtain an enriched sentence $\Phi'$ such that the problem $\Sat(\Phi')$ is polynomial-time equivalent to $\Sat(\Phi)$.
Replace every negated conjunct $\neg \phi_i$ of $\Phi$ with $\neg\phi_i' := \neg\left(\phi_i\wedge \bigwedge_{\rel{R}\in\tau}\rel{R}(\tuple{x}_\rel{R})\right)$, where each $\tuple{x}_\rel{R}$ is a tuple of new variables.
Observe that $\Phi'$ is not connected anymore.
However, $\Phi$ is more restrictive than $\Phi'$, as we enrich negated conjuncts of $\Phi$ with $\tau$-atoms.
That is, for every $\tau$-structure $\structure{A}$, $\structure{A}\models\Phi$ implies $\structure{A}\models\Phi'$.
Below, we check that $\Sat(\Phi)$ and $\Sat(\Phi')$ are polynomial-time equivalent.

\begin{lem}\label{lemma:non_empty_equiv1} 
For every connected $\GMSNP$ $\tau$-sentence $\Phi$ without redundant symbols there exists an enriched $\tau$-sentence $\Phi'$ such that $\Sat(\Phi)$ is polynomial-time reducible to $\Sat(\Phi')$.
\end{lem}
\proof
For every $\tau$-structure $\structure{A}$, let $\structure{A}'$ be the disjoint union $\structure{A} \uplus \left(\biguplus_{\rel{R}\in\tau}\structure{B}_\rel{R}\right)$.
Suppose that $\structure{A}\models\Phi$.
By Lemma~\ref{lemma:neg_r_case}, we assume that, for $\rel{R}$ in $\tau$, $\structure{B}_\rel{R}\models\Phi$.
Then, by Lemma~\ref{lem:snp_disjoint_union}, $\structure{A}'\models\Phi$, as $\structure{A}'$ is the disjoint union of structures satisfying $\Phi$.
But then, as $\Phi'$ is less restrictive than $\Phi$, we have $\structure{A}'\models\Phi'$.

Suppose that $\structure{A}'\models\Phi'$; then there exists a $\sigma$-expansion $\structure{A}'^\sigma$ that satisfies the first-order part $\phi'$ of $\Phi'$.
Define the $\sigma$-expansion $\structure{A}^\sigma$ as the substructure of $\structure{A}'^\sigma$ induced on $A$.
Suppose that there exist a negated conjunct $\neg\phi_i(\tuple{x}_i)$ of $\Phi$ and an assignment that makes it false in $\structure{A}^\sigma$.
But then we can take the negated conjunct $\neg\left(\phi_i\wedge \bigwedge_{\rel{R}\in\tau}\rel{R}(\tuple{x}_\rel{R})\right)$ of $\Phi'$, and assign $\tuple{b}_\rel{R}$ to each new tuple of variables $\tuple{x}_\rel{R}$.
Under this assignment the conjunct is false in $\structure{A}'^\sigma$, this is a contradiction.
\qed

\begin{lem}\label{lemma:non_empty_equiv2}
For every connected $\GMSNP$ $\tau$-sentence $\Phi$ without redundant symbols there exists an enriched $\tau$-sentence $\Phi'$ such that $\Sat(\Phi')$ is polynomial-time reducible to $\Sat(\Phi)$.
\end{lem}
\proof
Consider a $\tau$-structure $\structure{A}$ that is an input instance of $\Sat(\Phi')$.
Suppose that there exists $\rel{R}$ in $\tau$ such that $\rel{R}^\structure{A}=\varnothing$.
Then, $\structure{A}\models\Phi'$ because $\Phi'$ is enriched.
We reduce it to some fixed YES instance of $\Sat(\Phi)$.

Suppose that, for every $\rel{R}$ in $\tau$, $\rel{R}^\structure{A}\not=\varnothing$. Firstly, $\structure{A}\models\Phi$ implies $\structure{A}\models\Phi'$ because $\Phi$ is more restrictive than $\Phi'$.
Suppose that $\structure{A}\not\models\Phi$; then, for every $\sigma$-expansion $\structure{A}^\sigma$ of $\structure{A}$ there exists a negated conjunct $\neg\phi_i(\tuple{x}_i)$ of $\Phi$ and a tuple $\tuple{a}_i$ of elements of $\structure{A}$ such that $\structure{A}^\sigma\models\phi_i(\tuple{a}_i)$.
But then, take the negated conjunct $\neg\left(\phi_i(\tuple{x}_i)\wedge\bigwedge_{\rel{R}\in\tau}\rel{R}(\tuple{x}_\rel{R})\right)$, assign $\tuple{a}_i$ to $\tuple{x}_i$ and assign $\tuple{a}_\rel{R}$ to $\tuple{x}_\rel{R}$ such that, for every $\rel{R}$ in $\tau$, $\tuple{a}_\rel{R}\in\rel{R}^\structure{A}$.
At least one such $\tuple{a}_\rel{R}$ exists as $\rel{R}^\structure{A}\not=\varnothing$.
But then $\structure{A}^\sigma\models\left(\phi_i(\tuple{a}_i)\wedge\bigwedge_{\rel{R}\in\tau}\rel{R}(\tuple{a}_\rel{R})\right)$.
This implies that $\structure{A}\not\models\Phi'$.
\qed

\subsubsection*{Third stage: concatenation}\label{constr:concatenation} In this part, we construct a $\tau_1$-sentence $\Phi^1$ that is polynomial-time equivalent to $\Phi$.
By Lemmas~\ref{lemma:neg_r_case}, \ref{lemma:non_empty_equiv1}, and \ref{lemma:non_empty_equiv2}, we can assume that $\Phi$ is already enriched and has no redundant symbols.
Set $\tau_1 := \{\rel{P}\}$, where $\mathrm{arity}(\rel{P}) = \sum_{\rel{R}\in\tau}\mathrm{arity}(\rel{R})$. For two tuples $\tuple{x}=(x_1,\ldots,x_n),\tuple{y}=(y_1,\ldots,y_m)$, set $(\tuple{x},\tuple{y}):=(x_1,\ldots,x_n,y_1,\ldots,y_m)$.
The tuple $(\tuple{x}_1,\ldots,\tuple{x}_t)$ is defined similarly.

Let $\rel{R}_1,\ldots,\rel{R}_t$ be the relation symbols that constitute the signature $\tau$. For every negated conjunct $\neg\phi_i$ of $\Phi$, we transform it to a negated conjunct $\neg\phi_i^1$ of $\Phi^1$ as follows.
The $\sigma$-atoms of $\phi_i^1$ are the same as in $\phi_i$. For every $\tau$-atom $\rel{R}_j(\tuple{x})$ of $\phi_i$, add to $\phi_i^1$ a $\tau_1$-atom $\rel{P}(\tuple{y}_1,\ldots,\tuple{y}_t)$, where $\tuple{y}_j=\tuple{x}$ and all other variables of this atomic formula are new and they are used only there.
In the rest, we check the equivalence of the problems $\Sat(\Phi)$ and $\Sat(\Phi^1)$.

\begin{lem}\label{lemma:one_symbol1} 
$\Sat(\Phi)$ is polynomial-time reducible to $\Sat(\Phi^1)$.
\end{lem}
\proof
For every $\tau$-structure $\structure{A}$, we will construct a $\tau_1$-structure $\structure{A}_1$ such that $\structure{A}\models\Phi$ if and only if $\structure{A}_1\models\Phi^1$.
The structures have the same domain: $A_1=A$.
The relation $\rel{P}^{\structure{A}_1}$ is defined by the relations of $\tau$ as follows:
\begin{equation}\label{eq:construction_P_from_R}
\rel{P}^{\structure{A}_1}(\tuple{a}_1,\ldots,\tuple{a}_t)\longleftrightarrow\rel{R}_1^\structure{A}(\tuple{a}_1)\wedge\dots\wedge\rel{R}_t^\structure{A}(\tuple{a}_t).
\end{equation} 
Suppose that there exists $\rel{R}$ in $\tau$ such that $\rel{R}^\structure{A}=\varnothing$; then $\rel{P}^{\structure{A}_1}=\varnothing$ and $\structure{A}_1\models\Phi^1$.
But also $\structure{A}\models\Phi$, as $\Phi$ is enriched.
So, we can now assume that $\rel{R}^\structure{A}\not=\varnothing$ for every $\rel{R}$ in $\tau$.

Let $\structure{A}^\sigma$ and $\structure{A}_1^\sigma$ be two $\sigma$-expansions of $\structure{A}$ and $\structure{A}_1$ such that, for every $\rel{X}$ in $\sigma$, $\rel{X}^{\structure{A}^\sigma}=\rel{X}^{\structure{A}_1^\sigma}$.
It suffices to show that $\structure{A}^\sigma\models\forall\tuple{x}\;\phi(\tuple{x})$ if and only if $\structure{A}_1^\sigma\models\forall\tuple{x},\tuple{y}\;\phi^1(\tuple{x},\tuple{y})$, where $\tuple{y}$ represents variables that are added during the construction of $\Phi^1$.

Suppose that $\structure{A}^\sigma\not\models\forall\tuple{x}\;\phi(\tuple{x})$; then there exist a negated conjunct $\neg\phi_i(\tuple{x}_i)$ of $\Phi$ and a tuple $\tuple{a}_i\in A^{\tuple{x}_i}$ such that  $\structure{A}^\sigma\models\phi_i(\tuple{a}_i)$.
Take the $i$th negated conjunct $\neg\phi_i^1(\tuple{x}_i,\tuple{y}_i)$ of $\Phi^1$ and assign to $\tuple{x}_i$ exactly the same tuple $\tuple{a}_i\in A^{\tuple{x}_i}$.
As $\Phi$ is enriched, the conjunct $\phi_i(\tuple{x}_i)$ contains an atom $\rel{R}_j(\tuple{z}_j)$, for every $\rel{R}_j\in\tau$.
For every $j\in[t]$, let $\tuple{b}_j\in A^{\tuple{z}_j}$ be the restriction of $\tuple{a}_i$ on $\tuple{z}_j$.
Let $\rel{P}(\tuple{y}_1,\ldots,\tuple{y}_t)$ be a $\tau_1$-atom that contains a tuple of variables $\tuple{y}_j$ that is not contained in $\tuple{x}_i$ and is thus unassigned.
By construction, all the variables of $\tuple{y}_j$ appear in $\phi^1$ just once.
Therefore, we can assign the tuple $\tuple{b}_j$ to every such tuple $\tuple{y}_j$, for every $j\in[t]$.
Equation (\ref{eq:construction_P_from_R}) implies that $\structure{A}_1^\sigma\not\models\forall\tuple{x},\tuple{y}\;\phi(\tuple{x},\tuple{y})$. 

For the other direction, suppose that $\structure{A}_1^\sigma\not\models\forall\tuple{x},\tuple{y}\;\phi(\tuple{x},\tuple{y})$; then there exists a negated conjunct $\neg\phi_i^1(\tuple{x}_i,\tuple{y}_i)$ and tuples $\tuple{a}_i,\tuple{b}_i$ of elements of $A$ such that $\structure{A}_1^\sigma\models\phi_i^1(\tuple{a}_i,\tuple{b}_i)$.
Take the negated conjunct $\neg\phi_i(\tuple{x}_i)$ of $\Phi$ associated with $\neg\phi_i^1$ and assign $\tuple{a}_i$ to $\tuple{x}_i$.
We have $\structure{A}^\sigma\models\phi_i (\tuple{a}_i)$, by (\ref{eq:construction_P_from_R}).
\qed

\begin{lem}\label{lemma:one_symbol2} 
$\Sat(\Phi^1)$ is polynomial-time reducible to $\Sat(\Phi)$.
\end{lem}
\proof
For every $\tau_1$-structure $\structure{B}_1$, we construct a $\tau$-structure $\structure{B}$.
They have the same domain $B$, and every $\tau$-relation $\rel{R}_j^\structure{B}$ is defined as follows:
\begin{equation}\label{eq:construction_R_from_P}
\rel{R}_j^\structure{B}(\tuple{x}_j)\longleftrightarrow\exists\tuple{x}_1,\ldots,\tuple{x}_{j-1},\tuple{x}_{j+1},\ldots,\tuple{x}_t\;\rel{P}^{\structure{B}_1}(\tuple{x}_1,\ldots,\tuple{x}_t).
\end{equation}
Similarly as above, we consider two $\sigma$-expansions $\structure{B}^\sigma$ and $\structure{B}_1^\sigma$ such that, for all $\rel{X}$ in $\sigma$, $\rel{X}^{\structure{B}^\sigma}=\rel{X}^{\structure{B}_1^\sigma}$.

Suppose that $\structure{B}_1^\sigma\not\models\forall\tuple{x},\tuple{y}\;\phi^1(\tuple{x},\tuple{y})$; then there exists a negated conjunct $\neg\phi_i^1(\tuple{x}_i,\tuple{y}_i)$ and assignments $\tuple{a}_i\in B^{\tuple{x}_i}$ and $\tuple{b}_i\in B^{\tuple{y}_i}$ such that $\structure{B}_1^\sigma\models\phi_i^1(\tuple{a}_i,\tuple{b}_i)$.
Take the $i$th negated conjunct $\neg\phi_i(\tuple{x}_i)$ of $\Phi$ and assign $\tuple{a}_i$ to $\tuple{x}_i$.
As $\phi_i$ and $\phi_i^1$ have the same $\sigma$-part, $\structure{B}^\sigma$ satisfies every $\sigma$-atom and negated $\sigma$-atom of $\phi_i$.
Any $\tau$-atom of $\phi_i$ is satisfied, by (\ref{eq:construction_R_from_P}).
So $\structure{B}^\sigma\models\phi_i(\tuple{a}_i)$.

For the other direction, suppose that there exist a negated conjunct $\neg\phi_i(\tuple{x}_i)$ of $\Phi$ and and a tuple $\tuple{a}_i\in B^{\tuple{x}_i}$ such that $\structure{B}^\sigma\models\phi_i (\tuple{a}_i)$.
For every $\tau$-atom $\rel{R}_j(\tuple{z})$ of $\phi_i$, the conjunct $\phi_i^1$ contains a $\tau_1$-atom $\rel{P}(\tuple{y}_1,\ldots,\tuple{y}_t)$, where $\tuple{y}_j=\tuple{z}$.
Let $\tuple{c}\in B^{\tuple{z}}$ be the restriction of $\tuple{a}_i$ taken on $\tuple{z}$.
In particular, we have that $\structure{B}^\sigma\models\rel{R}_j(\tuple{c})$.
By (\ref{eq:construction_R_from_P}), for every $k\in[t]\setminus j$, there exists $\tuple{b}_k\in B^{\tuple{y}_k}$ such that $\structure{B}_1^\sigma\models\rel{P}(\tuple{b}_1,\ldots,\tuple{b}_t)$, where $\tuple{b}_j=\tuple{c}$.
So, for all $\tuple{y}_1,\ldots,\tuple{y}_t$ except for $\tuple{y}_j=\tuple{z}$, we assign $\tuple{b}_1,\ldots,\tuple{b}_t$ to them.
After doing this procedure for every $\tau$-atom of the conjunct $\phi_i^1$, all $\tau$-atoms of $\phi_i^1$ will be satisfied.
All $\sigma$-atoms and negated $\sigma$-atoms of $\phi_i$ are also satisfied because the $\sigma$-relations of $\structure{B}_1^\sigma$ are the same as in $\structure{B}^\sigma$.
This implies that $\structure{B}_1^\sigma\not\models\forall\tuple{x},\tuple{y}\;\phi^1(\tuple{x},\tuple{y})$ and we are done.
\qed

\proof[Proof of Theorem~\ref{th:gmsnp}]
By Lemma~\ref{lemma:gmsnp_connected}, we can assume that a given GMSNP $\tau$-sentence $\Phi$ is connected.
By Lemma~\ref{lemma:neg_r_case}, we can assume that $\Phi$ has no redundant symbols.
Let $\Phi'$ be constructed from $\Phi$ as in the \emph{second stage}.
By Lemmas~\ref{lemma:non_empty_equiv1}~and~\ref{lemma:non_empty_equiv2}, $\Sat(\Phi)$ and $\Sat(\Phi')$ are polynomial-time equivalent.
Let $\Phi^1$ be constructed from $\Phi'$ as in the \emph{third stage}.
By Lemmas~\ref{lemma:one_symbol1}~and~\ref{lemma:one_symbol2}, $\Sat(\Phi')$ and $\Sat(\Phi^1)$ are polynomial-time equivalent.
\qed

%%%%%%%%%%%%%%%%%%%%%%%%%%%%%%%%%%%%%%%%%%%%%%%%%%%%%%%%%%%%%%%%%%%%%%%%%%%%%%%%
\subsection{Matrix Partitions}\label{section:mp}

In this subsection, we define the classes $\GMPARTineq$ and $\MPART$ and prove Theorem~\ref{th:mpart}.

\begin{figure}
\centering
\includegraphics[width=0.55\textwidth]{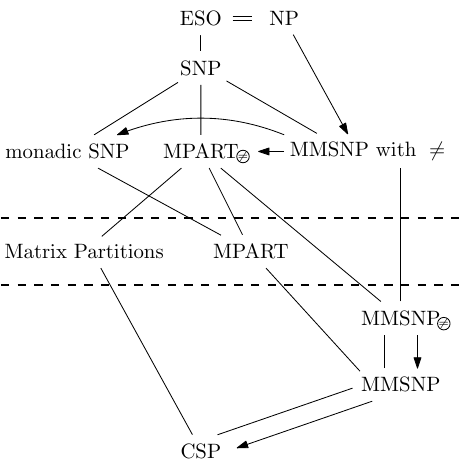}
\caption{The classes introduced in Subsection~\ref{section:mp} and their relation to other classes from this article. Dashed lines divide the figure in 3 parts: ``dichotomy'', ``unknown'', ``no dichotomy''. Undirected edges stand for inclusions and directed edges stand for inclusions under polynomial-time reductions.}
\label{fig:mpart}
\end{figure}

\begin{defi}\label{def:GMPART}
An $\SNP$ sentence $\Phi$ in standard form is in \emph{$\GMPARTineq$} if
\begin{inparaenum}[(a)]
\item each $\alpha_i$ is either a conjunction only of non-negated $\tau$-atoms or a conjunction only of negated $\tau$-atoms,
\item each $\beta_i$ is a conjunction of unary $\sigma$-atoms or negated unary $\sigma$-atoms, and
\item for every inequality $x_j\not=x_j'$ of each $\epsilon_i$ there exists a $\tau$-atom or a negated $\tau$-atom in $\alpha_i$ that contains both $x_j$ and $x_j'$.
\end{inparaenum}
\end{defi}
The class $\GMPARTineq$ contains both Matrix Partition problems and $\GMMSNPineq$ from Subsection~\ref{section:gmmsnpineq} but it is not contained in monadic $\SNP$ without inequality as $\GMPARTineq$ allows inequalities.
Denote by \emph{$\MPART$} the fragment of $\GMPARTineq$ where each $\epsilon_i$ is empty.
Therefore, it will be contained in the gap between $\MMSNP$ and monadic SNP without inequality, see \Cref{fig:mpart}.

\begin{thm}\label{th:mpart}
The class $\MMSNP$ is strictly contained in $\MPART$, and $\MPART$ is strictly contained in $\GMPARTineq$.
Moreover, for every problem in $\MMSNPineq$ there exists a polynomial-time equivalent problem in $\GMPARTineq$.
\end{thm}
\proof
The containment in both cases follows from the definition.
Below, we show, that the containment is strict.

Let $\Phi\in\MPART$ be a sentence that accepts only complete directed graphs with loops: 
\begin{equation*}
\forall x,y\;\neg\bigl(\neg\rel{E}(x,y)\bigr).
\end{equation*}
As $\Sat(\Phi)$ is not closed under inverse homomorphisms, $\Phi$ is not expressible in $\MMSNP$, by Theorem~\ref{th:monotone-hom-closed}.

Let $\Phi\in\GMPARTineq$ be a sentence that accepts only an isolated vertex or an isolated vertex with a loop:
\begin{equation*}
\forall x,y\; \neg\bigl(\rel{E}(x,y)\wedge x\not=y\bigr) \wedge \neg\bigl(\neg\rel{E}(x,y)\wedge x\not=y\bigr)
\end{equation*}

Suppose that a graph $\structure{G}$ (with loops allowed) satisfies a sentence $\Psi$ in $\MPART$.
Take any vertex $v$ of $\structure{G}$ and add to $\structure{G}$ a twin $v'$ of $v$, i.e., the sets of vertices adjacent to $v$ and to $v'$ are the same.
The resulting graph $\structure{G}'$ still satisfies $\Psi$ as one can choose the same $\sigma$-relations for $v'$ as were chosen for $v$ in the $\sigma$-expansion $\structure{G}^\sigma$ witnessing $\structure{G}\models\Psi$.
Let $(\structure{G}')^\sigma$ be this $\sigma$-expansion.
If $(\structure{G}')^\sigma$ does not satisfy the first-order part of $\Psi$, then, for some negated conjunct $\neg\psi(\tuple{x})$ of $\Psi$ and for some $\tuple{g}\in (G')^{\tuple{x}}$, we have $(\structure{G}')^\sigma\models\psi(\tuple{g})$.
Then, as $v$ and $v'$ are twins, we have $\structure{G}^\sigma\models\psi(\tuple{g}_{v'\to v})$, where the tuple $\tuple{g}_{v'\to v}$ is obtained from $\tuple{g}$ by replacing each appearance of $v'$ with $v$.
In particular, if $\Psi$ holds in the graph consisting of an isolated vertex, it will hold in the graph consisting of two isolated vertices.
Therefore, $\Phi$ is not expressible in $\MPART$.

Now, we prove that every problem in $\MMSNPineq$ is polynomial-time equivalent to a problem in $\GMPARTineq$ which implies no dichotomy for the latter class.
Let $\Phi$ be of some signature $\tau$. 
Add to $\tau$ a new binary relation symbol $\rel{N}$. 
The $\tau\cup\{\rel{N}\}$-sentence $\Psi$ is constructed as follows.
\begin{inparaenum}[(a)]
\item Replace every atomic formula of $\Phi$ of the form $x\not=y$ with $\rel{N}(x,y)$.
\item Add to the first-order part the two following negated conjuncts:
\begin{equation*}
\forall x,y\; \neg\rel{N}(x,x)\wedge\neg\bigl(\neg\rel{N}(x,y)\wedge x\not=y\bigr).
\end{equation*}
Let $\structure{A}$ be a $\tau$-structure and $\structure{A}'$ be its $\{\rel{N}\}$-expansion, where $\rel{N}$ is interpreted as the inequality relation. 
Then, associating $\structure{A}'$ with $\structure{A}$ we reduce $\Phi$ to $\Psi$. 
For the other direction, if $\rel{N}$ is not interpreted as the inequality in the input $\tau\cup\{\rel{N}\}$-structure, then $\Psi$ rejects it, so we reduce it to some fixed NO instance of $\Phi$.
Otherwise, we reduce it to the $\tau$-reduct of the input instance.
\end{inparaenum}
\qed

Although $\GMPARTineq$ is related to $\MPART$ in the same way as $\GMMSNPineq$ is related to $\MMSNP$, the idea of the proof of Theorem~\ref{th:gmmsnpineq_embeds_mmsnp} do not transfer to them because now negated conjuncts may contain negated $\tau$-atoms. 
This gives motivation to ask the following question.
\begin{qu}
Does the class $\MPART$ have a dichotomy?
\end{qu}

%%%%%%%%%%%%%%%%%%%%%%%%%%%%%%%%%%%%%%%%%%%%%%%%%%%%%%%%%%%%%%%%%%%%%%%%%%%%%%%%%%%%%
\section{NP is polynomial-time equivalent to MMSNP with inequality}\label{section:snp_subclasses}

In this section, we provide a detailed proof of the following theorem.

\begin{thmC}[\cite{federvardi1998}]\label{th:np_subset_mmsnpineq}
For every problem $\Omega$ in $\NP$ there exists a polynomial-time equivalent sentence $\Phi$ in $\MMSNPineq$.
\end{thmC}

The signature of the sentence $\Phi$ is denoted with $\tau$.
We will construct two mappings: \begin{inparaenum}[(i)] \item $\func{r_{easy}}$ from the set of input strings of $\Omega$ to the set of $\tau$-structures and \item $\func{r_{hard}}$ in the opposite direction, \end{inparaenum}
such that 
\begin{compactitem}
  \item for every input string $\tuple{s}$ of $\Omega$, $\tuple{s}$ is accepted by $\Omega$ if and only if $\func{r_{easy}}(\tuple{s})\models\Phi$, and 
  \item for every $\tau$-structure $\structure{A}$, we have $\structure{A}\models\Phi$ if and only if $\func{r_{hard}}(\structure{A})$ is accepted by $\Omega$.
\end{compactitem}

First, without loss of generality we assume that $\Omega$ is decided by an \emph{oblivious} Turing machine $M$. 
This property means that the head movement depends only on the size of the input string.  
We build a sentence $\Phi$ in $\MMSNPineq$ that shall operate on an input $\structure{A}$ which ought to describe the space-time diagram of an input string $\tuple{x}$ of $M$. 
We use two binary input predicates $\rel{succ}$ and $\rel{next}$ for space and time respectively, together with additional predicates to obtain a grid like structure.

Obliviousness allows us to check the computation with monadic predicates only, and it is not hard to prove that there exists a polynomial-time reduction $\func{r_{easy}}$ from $\Omega$ to $\Sat(\Phi)$.
The difficult part is the converse direction, where one has to deal appropriately with structures that are degenerate in the sense that they are not in the image of the reduction $\func{r_{easy}}$.
It is precisely this part of the proof that is fully absent in~\cite{federvardi1998}. 
Our construction of $\Phi$ ensures the following.
\begin{compactitem}
\item If $\structure{A}$ is \emph{over-complete} (\ie, contains too many relational tuples), then $\structure{A}\not\models\Phi$. Here, $\structure{A}$ does not satisfy some first-order properties that are defined using ``inequality''. This is the case, where the expressive power of ``inequality'' is used.
\item If $\structure{A}$ is \emph{incomplete} either locally -- it is like a grid with holes -- or globally -- too small to simulate a complete run of $M$, then $\structure{A}\models\Phi$. This is achieved by a monadic marking scheme of $\structure{A}$ by the sentence $\Phi$ and a relativisation of the verification of $M$'s computation to the marked part.
\item If $\structure{A}$ is in an \emph{appropriate form} and can simulate a run of $M$, then $\structure{A}\models\Phi$ if and only if $M$ accepts.
\end{compactitem}

\subsection*{Convention on Turing machines}
Every Turing machine that we further consider is nondeterministic. We follow the convention of~\cite{papadimitriou_book} and assume that the input tape is one-way infinite. The alphabet is a set $\Sigma\cup\{\triangleright,\sqcup\}$, where $\triangleright$ is the \emph{first} symbol marking the left end of the tape, $\sqcup$ is the \emph{blank} symbol, and $\Sigma$ contains all other symbols and may vary for different machines. The set of states of the machine is denoted by $\family{Q}$. The machine input is a string of the form $\triangleright(\Sigma)^n\sqcup^\infty$ for $n\in\mathbb{N}$, \ie, the first symbol is followed by a string of symbols from $\Sigma$ of length $n$, and the rest consists of blank symbols. The number $n$ of elements from $\Sigma$ in the input string is called the \emph{size} of the input. The cell  that contains $\triangleright$ is called the \emph{first cell}. If the head arrives in the first cell during the execution, then it will write $\triangleright$ and move to the right. Before the execution starts, the head of the machine is at the first cell. We assume that every machine $M$ halts in polynomial time in the size of the input. That is, for every machine $M$ there exists $k>0$ and a function $\func{f}\colon\mathbb{N}\to\mathbb{N}$ such that $\func{f}(n)\in O(n^k)$ and, for every input of size $n$, the machine $M$ halts in at most $\func{f}(n)$ steps. Finally, we assume that every machine $M$ that we consider is not trivial, \ie, there exists at least one input string which is accepted and at least one which is rejected. 
We fix a string $\tuple{s}_{\mathrm{Y}}$ which is accepted by $M$ and a string $\tuple{s}_{\mathrm{N}}$ which is rejected.
\subsection{Obliviousness}\label{subsection:obliviousness}

A Turing machine $M$ is called \emph{oblivious} if there exists a function $\func{f}\colon\mathbb{N}\times\mathbb{N}\to\mathbb{N}$ such that, for every input $\tuple{x}$ of size $\card{\tuple{x}}=n$, at any moment of time $t\in\mathbb{N}$, the head of the machine is at the $\func{f}(n,t)$th cell. Oblivious Turing machines are introduced by Pippenger and Fischer in~\cite{pippenger1979}.

In order to ensure that an input instance $\structure{A}$ of a sentence in $\MMSNPineq$ can simulate the whole execution of $M$, we need to know
\begin{inparaenum}[(a)] \item the position of its head at any moment of time and \item the precise time when the machine will halt.
\end{inparaenum}

In the $\MMSNPineq$ sentence which is supposed to simulate the oblivious Turing machine, the head movement will be described by an existentially defined relation.
In order to make the reductions $\func{r_{easy}}$ and $\func{r_{hard}}$ universal, we assume that the head movement in every oblivious Turing machine is the same and similar to the one from Theorem 3 in~\cite{pippenger1979}.
That is, for every input of size $n$, the head first does $n+1$ steps to the right until it reaches a cell with the blank symbol, then returns back to the first cell, then does $n+2$ steps to the right, returns to the first cell again, and so on. 
When the machine halts, the head is at the first cell. 
Call an oblivious machine \emph{back-and-forth} if its head movement is as described above. 
Say that the head of a back-and-forth machine does a \emph{round trip} if it starts at the first cell, goes right until it scans a blank symbol, and returns back to the first cell. 

The ideas from the proof of Theorem 3 in~\cite{pippenger1979} imply the following statement.

\begin{lem}\label{lem:pippenger}
For every machine $M$ there exists an equivalent machine $M'$ which is back-and-forth.
\end{lem}
\proof
For every transition of the machine $M$, the back-and-forth machine $M'$ spends one round trip. Those transitions of $M$ where the head goes from the $i$th cell to the $(i+1)$th cell are simulated when the head of $M'$ is going from left to right. The transitions of $M$ where the head goes from the $i$th cell to the $(i-1)$th cell are simulated when the head of $M'$ is going in the opposite direction.
\qed

\medskip

Say that a back-and-forth machine is \emph{precise} if there exist $k\in\mathbb{N}$ and $\func{g}(n)\in O(n^k)$ such that, for every input string of length $n$, the head does precisely $\func{g}(n)$ round trips as above before it halts. 
If the value $\func{g}(n)$ is known, then one can obtain the exact halting time $\func{f}(n)$ as follows.
To do the first round trip on an input of size $n$, the head will need to make $(n+1)+(n+1)$ steps.
For each next round trip, the number of steps will increase by 2, so the last round trip is done in $(n+\func{g}(n))+(n+\func{g}(n))$ steps.
Therefore, the total execution time would be
\begin{equation*}
\func{f}(n) := \func{g}(n)\bigl(\func{g}(n)+2n+1\bigr)
\end{equation*}

The preciseness condition is necessary for us because we need to know whether a given input space-time diagram is large enough to simulate the execution of the machine. The goal of Subsection~\ref{subsection:obliviousness} is to prove the following.
\begin{thm}\label{th:precise}
For every machine $M$ there exists a polynomial-time equivalent machine $M'$ which is precise.
\end{thm}
\proof
By Lemma~\ref{lem:pippenger}, we may assume that $M$ is back-and-forth. The goal is to construct a precise machine $M'$ that simulates $M$. 

Recall that, by convention, there exist $k>0$ and $\func{f}\in O(n^k)$ such that, for every input of size $n$, the machine $M$ halts in at most $\func{f}(n)$ steps. Therefore, there exists $\func{g}\in O(n^k)$ such that $M$ does at most $\func{g}(n)$ round trips. 
We may assume without loss of generality that $\func{g}(n)=cn^k$, where $c,k\in\mathbb{N}$. 
It suffices to modify $M$ so that it does precisely $\func{g}'(n)$ round trips, where $\func{g}'(n)\in O(n^k)$.
This can be done by adding some procedure that has running time a little bit greater than $cn^k$ and that requires the same amount of time for all input strings of the same size.

Recall that the input string of $M$ is of the form $\triangleright(\Sigma)^n\sqcup^\infty$. 
We want it to be of the form $\triangleright0^{m}(\Sigma)^n\sqcup^\infty$. 
That is, we want to add a string consisting of $0$s of length $m$ between the first cell and the input string so that the machine $M'$ would write on this string all numbers from $0$ to $2^{m}$ in binary encoding.
We want $m$ to be at least $\lceil\log_2 cn^k\rceil$ and that $2^m\in O(n^k)$.
We call this process the \emph{binary enumeration}.

The alphabet of the new machine $M'$ is $\{\triangleright,\sqcup,0,1\}\cup(\Sigma\times\{0,1\})$, where $\{\triangleright,\sqcup\}\cup\Sigma$ is the alphabet of $M$. 
At the first round trip, the machine makes the second coordinate equal to the parity of the cell, \ie, for every $i\in[n]$, if a cell contains a symbol $(s,t)$, $s\in\Sigma$, $t\in\{0,1\}$, then it writes $(s,i \mod 2)$ instead. 
After the first round trip, the string of the second coordinate symbols has the form $(01)^{n/2}$. 
Then, at each next round trip, the machine $M'$ decides whether it preserves or changes the second-coordinate symbol depending on one of the two states: $Q_+, Q_-$, which are defined as follows.
The machine $M'$ starts a round trip having the state $Q_+$.
If, right after reading ``0'', it reads ``1'', $M'$ changes to the state $Q_-$ until it reads ``01'' for the next time.
If $M'$ has the state $Q_-$, then it rewrites ``0'' with ``1'' and vice versa, otherwise, it preserves the second coordinate symbol.
Therefore, after the second round trip, the second coordinate of the string will have the form $(0011)^{n/4}$, after the third, it will look like $(00001111)^{n/8}$, and so on. 
The procedure runs until the second coordinate string has the form $0^n$. 
It requires $\lceil\log_2 n\rceil + 1$ round trips. 
Moreover, at each round trip, the head must move the input string one position to the right. 
This can be done during the movement from left to right: the machine reads the symbol $(s_i,t_i)$ from the $i$th cell, remembers it, writes the previously scanned symbol $(s_{i-1},t_{i-1})$, moves to the right, scans $(s_{i+1},t_{i+1})$ and writes $(s_i,t_i)$ instead, and so on.

If we repeat this procedure $k$ times, then the machine $M'$ will make $k\lceil\log_2 n\rceil + k$ round trips.
There exists some constant $c'\in\mathbb{N}$  such that $c'$ depends only on $k$ and $c$ and that $k\lceil\log_2 n\rceil + k +c' \geq \lceil\log_2 cn^k\rceil$.
Notice that $2^{k\lceil\log_2 n\rceil + k +c'}\in O(n^k)$, so we can put $m$ to be equal to $k\lceil\log_2 n\rceil + k +c'$.
As $c'$ depends only on $k$ and $c$, we can move the first coordinate string to the right by $c'$ cells.

Now we can assume that the input string looks like $\triangleright0^{m}(\Sigma)^n\sqcup^\infty$. The machine $M'$ can simulate the movement of the head of $M$ and do the binary enumeration during the same round trip. By our assumption, the simulation of $M$ finishes earlier than the binary enumeration. Therefore, we require the head of $M'$ to repeat round trips until the binary enumeration is finished in order to keep it back-and-forth. The binary enumeration is done in $O(n^k)$ steps. The number of steps depends only on the size of the input, so the resulting machine is precise.
\qed

\medskip

\subsection{Construction of the space-time diagram.}\label{subsection:space_time_diagram}

By Theorem~\ref{th:precise}, we can assume that the machine $M$ is precise. Therefore, we can fix the functions $\func{f}(n)$ and $\func{g}(n)$ such that $M$ halts precisely in time $\func{f}(n)$ and does precisely $\func{g}(n)$ round trips for every input of size $n$, where
\begin{equation*}
\func{f}(n) := \func{g}(n)\bigl(\func{g}(n)+2n+1\bigr)
\end{equation*}

Let $\tuple{s}\in\Sigma^n$ be an input string of $M$. In this section, we describe the relational structure $\structure{A}$ such that $\func{r_{easy}}(\tuple{s}) = \structure{A}$. 
The structure $\structure{A}$ will have a form of a space-time diagram, where the horizontal axis represents the string of symbols, \ie, \emph{space}, and the vertical axis represents \emph{time}.

\subsubsection*{Domain of $\structure{A}$}
The domain $A$ of $\structure{A}$ is the following disjoint union:
\begin{equation}
A = A_0^\leftarrow\uplus \biguplus_{i=1}^{\func{g}(n)}(A_i^\rightarrow\uplus A_i^\leftarrow) \uplus A_{\func{g}(n)+1}^\rightarrow
\end{equation}
Here, the set $A_0^\leftarrow$ describes the machine tape before the execution starts.
For every $i\in[\func{g}(n)]$, the set $A_i^\rightarrow$ represents the machine tape during the time when the head goes from the first cell to the end of the input string while the set $A_i^\leftarrow$ does the same for the movement in the opposite direction.

\begin{figure}
\begin{subfigure}{0.49\textwidth}
\centering
\includegraphics[width=\textwidth]{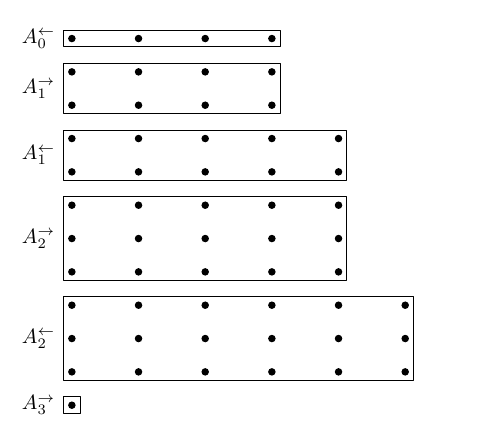}
\caption{The domain $A$ of $\structure{A}$.}
\label{fig:domain}
\end{subfigure}
\begin{subfigure}{0.49\textwidth}
\centering
\includegraphics[width=\textwidth]{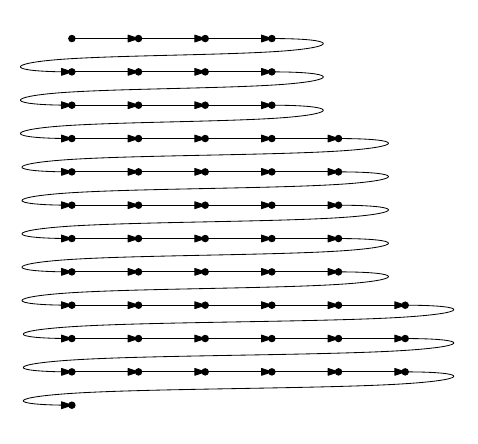}
\caption{The relation $\rel{succ}^\structure{A}$.}
\label{fig:succ}
\end{subfigure}\\
\begin{subfigure}{0.49\textwidth}
\centering
\includegraphics[width=\textwidth]{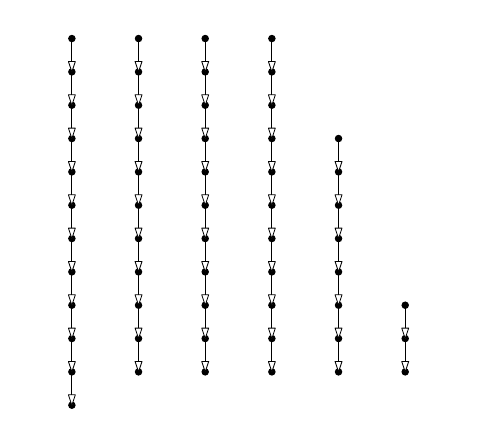}
\caption{The relation $\rel{next}^\structure{A}$.}
\label{fig:next}
\end{subfigure}
\begin{subfigure}{0.49\textwidth}
\centering
\includegraphics[width=\textwidth]{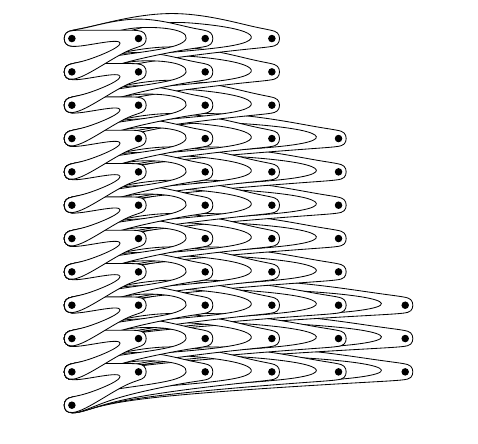}
\caption{The relation $\rel{row}^\structure{A}$.}
\label{fig:row}
\end{subfigure}
\begin{subfigure}{0.35\textwidth}
\centering
\includegraphics[width=\textwidth]{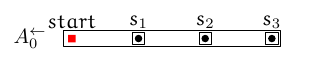}
\caption{The relations $\rel{start}^\structure{A}$ and $\rel{s}_1^\structure{A},\rel{s}_2^\structure{A},\rel{s}_3^\structure{A}$.}
\label{fig:start_symbols}
\end{subfigure}
\caption{The domain and the relations for a structure $\structure{A}$ from Example~\ref{ex:construction}.}
\label{fig:construction}
\end{figure}

We put $A_0^\leftarrow = [n+1]\times[1]$.
For every $i\in[\func{g}(n)]$, we put $A_i^\rightarrow = [n+i]\times[n+i-2]$ and $A_i^\leftarrow = [n+i+1]\times[n+i-2]$, where the first coordinate denotes the number of columns and the second coordinate denotes the number of rows.
Finally, $A_{\func{g}(n)+1}^\rightarrow = [1]\times[1]$ is a one-element set.
See \Cref{fig:m_to_phi_example} and \Cref{fig:domain} for an illustration.

Notice, that the total number of rows in this grid is equal to
\begin{equation*}
1 + 1 + \sum_{i=1}^{\func{g}(n)} 2(n+i-2) = \bigl(\func{g}(n)\bigr)^2 + (2n-3)\func{g}(n) + 2
\end{equation*}

For every set of $\{A_0^\leftarrow,A_1^\rightarrow,A_1^\leftarrow,\dots,A_{\func{g}(n)+1}^\rightarrow\}$, let $\func{l}$ be the function that outputs the maximal value of the first component, that is,
\begin{equation*}
\func{l}(A_0^\leftarrow)=n+1\;\mathrm{and},\; \mathrm{for}\; i\in[\func{g}(n)],\; \func{l}(A_i^\rightarrow) = n+i,\; \func{l}(A_i^\leftarrow) = n+i+1\;\mathrm{and}\;\func{l}\bigl(A_{\func{g}(n)+1}^\rightarrow\bigr) = 1
\end{equation*}
Similarly, define the function $\func{h}$ that outputs the maximal value of the second component:
\begin{equation*}
\func{h}(A_0^\leftarrow)=1\;\mathrm{and},\; \mathrm{for}\; i\in[\func{g}(n)],\; \func{h}(A_i^\rightarrow) = \func{h}(A_i^\leftarrow) = n+i-2\;\mathrm{and}\;\func{h}\bigl(A_{\func{g}(n)+1}^\rightarrow\bigr) = 1
\end{equation*}

\subsubsection*{Relations of $\structure{A}$}
The structure $\structure{A}$ has the following relational signature:
\begin{equation*}
\tau = \bigl\{\rel{s}(\cdot)\mid s\in\Sigma\setminus\{\sqcup\}\bigr\}\uplus\bigl\{\rel{start}(\cdot), \rel{succ}(\cdot,\cdot), \rel{next}(\cdot,\cdot),\rel{row}(\cdot,\cdot,\cdot)\bigr\}
\end{equation*}
\begin{compactitem}
\item The binary relation $\rel{succ}^\structure{A}$ denotes horizontal arcs.
It links two neighbours in the same row and also it connects the last element of a row to the first element of the next row. 
Everywhere in this section, it is denoted by black-headed arcs.
Formally, a pair $(\tuple{x},\tuple{y})\in A^2$ is added to $\rel{succ}^\structure{A}$ if and only if there exists $i\in\{0,\ldots,\func{g}(n)\}$ such that one of the following holds:
\begin{compactitem}
\item $\tuple{x},\tuple{y}\in A_i^\rightarrow$ and $x_1+1=y_1$ and $x_2=y_2$;
\item $\tuple{x},\tuple{y}\in A_i^\leftarrow$ and $x_1+1=y_1$ and $x_2=y_2$;
\item $\tuple{x},\tuple{y}\in A_i^\rightarrow$ and $x_1=\func{l}(A_i^\rightarrow)$ and $y_1=1$ and $x_2+1=y_2$;
\item $\tuple{x},\tuple{y}\in A_i^\leftarrow$ and $x_1=\func{l}(A_i^\leftarrow)$ and $y_1=1$ and $x_2+1=y_2$;
\item $\tuple{x}\in A_i^\rightarrow$ and $\tuple{y}\in A_i^\leftarrow$ and $\tuple{x} = (\func{l}(A_i^\rightarrow), \func{h}(A_i^\rightarrow))$ and $\tuple{y} = (1, 1)$;
\item $\tuple{x}\in A_i^\leftarrow$ and $\tuple{y}\in A_{i+1}^\rightarrow$ and $\tuple{x} = (\func{l}(A_i^\leftarrow), \func{h}(A_i^\leftarrow))$ and $\tuple{y} = (1, 1)$.
\end{compactitem}
\item The binary relation $\rel{next}^\structure{A}$ denotes vertical arcs.
It links two neighbours in the same column and is denoted by white-headed arcs elsewhere.
Formally, a pair $(\tuple{x},\tuple{y})\in A^2$ is added to $\rel{next}^\structure{A}$ if and only if there exists $i\in\{0,\ldots,\func{g}(n)\}$ such that one of the following holds:
\begin{compactitem}
\item $\tuple{x},\tuple{y}\in A_i^\rightarrow$ and $x_1=y_1$ and $x_2+1=y_2$;
\item $\tuple{x},\tuple{y}\in A_i^\leftarrow$ and $x_1=y_1$ and $x_2+1=y_2$;
\item $\tuple{x}\in A_i^\rightarrow$ and $\tuple{y}\in A_i^\leftarrow$ and $x_1=y_1$ and $x_2 = \func{h}(A_i^\rightarrow)$ and $y_2=1$;
\item $\tuple{x}\in A_i^\leftarrow$ and $\tuple{y}\in A_{i+1}^\rightarrow$ and $x_1=y_1$ and $x_2 = \func{h}(A_i^\leftarrow)$ and $y_2=1$.
\end{compactitem}
\item The relation $\rel{row}^\structure{A}$ is ternary. It describes all the triples, where the first and the third coordinates are the leftmost elements of two consecutive rows while the second coordinate belongs to the same row as the first one.
Formally, a triple $(\tuple{x},\tuple{y},\tuple{z})\in A^3$ is added to $\rel{row}^\structure{A}$ if and only if the following holds:
\begin{center}
$x_1=1$ \: and \: $\rel{next}(\tuple{x},\tuple{z})$ \: and \: $y_1>1$ \: and \: $x_2=y_2$.
\end{center}

\item The unary relation $\rel{start}^\structure{A}$ represents the point of reference of the diagram. 
there exists a unique element in $\rel{start}^\structure{A}$ -- the first element of the first row. It is denoted by a red square everywhere in this section. 
Formally, an element $\tuple{x}\in A$ belongs to $\rel{start}^\structure{A}$ if and only if $\tuple{x}\in A_0^\leftarrow$ and $\tuple{x} = (1,1)$.
\item For every symbol $s\in\Sigma$, the unary relation $\rel{s}^\structure{A}$ captures all the cells of the input string $\tuple{s}$ which contain the symbol $s$.
Formally, an element $\tuple{x}\in A$ belongs to $\rel{s}^\structure{A}$ if and only if $\tuple{x}\in A_0^\leftarrow$ and the $(x_1-1)$th cell of $\tuple{s}$ contains the symbol $s$.
\end{compactitem}

\begin{exa}\label{ex:construction}
Suppose that an input string $\tuple{s} = (s_1s_2s_3)$ of the machine $M$ has size $3$ and that $\func{g}(3)=2$.
This means, that the machine $M$ has to make two round trips and then halt.
Then the domain of the $\tau$-structure $\structure{A} = \func{r_{easy}}(\tuple{s})$ is displayed on \Cref{fig:domain} and is the following disjoint union
\begin{equation*}
([4]\times[1])\uplus([4]\times[2])\uplus([5]\times[2])\uplus([5]\times[3])\uplus([6]\times[3])\uplus([1]\times[1])
\end{equation*}
The relations $\rel{succ}^\structure{A}$, $\rel{next}^\structure{A}$, and $\rel{row}^\structure{A}$ are exactly as on \Cref{fig:succ,fig:next,fig:row} respectively.
The \Cref{fig:start_symbols} displays the relation $\rel{start}^\structure{A}$ and the relations $\rel{s}_1^\structure{A},\rel{s}_2^\structure{A},\rel{s}_3^\structure{A}$ which represent the symbols of the string $\tuple{s}$.
\end{exa}

\subsection{Construction of the sentence.}\label{subsection:construction_of_phi}
The existential relations of $\Phi$ in $\MMSNPineq$ are as follows:
\begin{equation*}
\sigma = \bigl\{ \rel{I}(\cdot),\rel{M}(\cdot),\rel{H}(\cdot)\bigr\}\uplus \bigl\{\rel{S}(\cdot)\mid s\in\Sigma\bigr\}\uplus\bigl\{\rel{Q}(\cdot)\mid q\in\family{Q}\bigr\}
\end{equation*} 
\begin{compactitem}
\item $\rel{I}$ (\emph{initial}) contains the elements of the first row, it is used in order to generate all $\rel{M}$ elements and to assign the initial symbols: $\rel{s}(x)\to\rel{S}(x)$. %It is denoted by 
\item $\rel{M}$ (\emph{marked}) determines the region, where the head movement is mandatory. %On figures, it is denoted by small white disks.small white disks as well as $\rel{M}$.
\item $\rel{H}$ (\emph{head}) contains all the positions of $M$'s head at every moment of time. %It is denoted by large white disks.
\item Relations $\rel{S}$ describe the string symbols at any moment of time.
\item Relations $\rel{Q}$ represent the current state of the machine.
\end{compactitem}

\Cref{subsection:construction_of_phi} is divided into several paragraphs. 
Each of them describes a certain subroutine of the simulation of $M$.

\subsubsection*{Initiating}

Here, we explain how to find the part of the input from which we later construct the input string of $M$ and how the simulation of $M$ is supposed to start.

\begin{figure}
\centering
\includegraphics[width=0.75\textwidth]{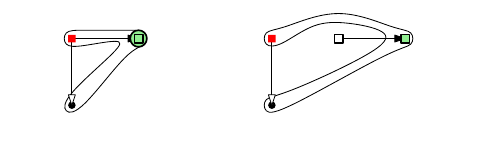}
\caption{The rule (left) initiating the relations $\rel{I}$ and $\rel{H}$ (denoted by $\Box$ and by a circle, respectively) and the rule (right) that forces the relation $\rel{I}$ to spread. The $\rel{I}$- and $\rel{H}$-atoms in the head of each rule are coloured in green.}
\label{fig:marked_base}
\end{figure}

The relations $\rel{I}$ and $\rel{H}$ are initiated by the $\rel{succ}$-neighbour of the $\rel{start}$ if the structure induced on the $\rel{start}$-element and its $\rel{succ}$- and $\rel{next}$-neighbours repeats the grid-structure from \Cref{subsection:space_time_diagram}.
Then, the relation $\rel{I}$ is forced to spread  along the $\rel{succ}$-arcs either until the end of the first row or until an element that is not contained in a relation $\rel{s}$ for $s\in\Sigma$.
See (\ref{eq:initial1}), (\ref{eq:initial2}), and Figure~\ref{fig:marked_base} for an illustration.
Although, in \cref{eq:initial1}, the head contains a conjunction of two atomic formulas which is not allowed by $\MMSNPineq$, this clause can be transformed to a conjunction of allowed clauses as follows:
\begin{equation*}
A\rightarrow(B\wedge C)\equiv (A\rightarrow B)\wedge (A\rightarrow C)\equiv \neg (A\wedge \neg B)\wedge \neg(A\wedge\neg C)
\end{equation*}

\begin{equation}\label{eq:initial1}
\bigwedge_{s\in\Sigma}\bigl(\rel{start}(x)\wedge\rel{next}(x,z)\wedge\rel{succ}(x,y)\wedge\rel{row}(x,y,z)\wedge\rel{s}(y)\rightarrow\rel{I}(y)\wedge\rel{H}(y)\bigr)
\end{equation}
\begin{equation}\label{eq:initial2}
\bigwedge_{s\in\Sigma}\bigl(\rel{start}(w)\wedge\rel{next}(w,z)\wedge\rel{I}(x)\wedge\rel{succ}(x,y)\wedge\rel{row}(w,y,z)\wedge\rel{s}(y)\rightarrow\rel{I}(y)\bigr)
\end{equation}

If the machine $M$ is at the state $q_0\in\family{Q}$ at the start of the execution, then every element in $\rel{I}$ must be also in $\rel{Q}_0$:
\begin{equation}\label{eq:initiating_q}
\rel{I}(x)\to\rel{Q}_0(x)
\end{equation}
If an element is forced to be in $\rel{I}$, then it is contained in some input relation $\rel{s}$.
In this case, it is also forced to be in the existential relation $\rel{S}$ associated with $\rel{s}$:
\begin{equation}\label{eq:initiating_s}
\bigwedge_{s\in\Sigma}\bigl(\rel{I}(x)\wedge\rel{s}(x)\to\rel{S}(x)\bigr)
\end{equation}

\subsubsection*{Determining the substructure where the simulation of $M$ happens}
During the reduction $\func{r}_{\mathrm{hard}}$, we determine in polynomial-time, whether the input structure is sufficiently large to simulate the execution of $M$.
For this reason, we have the relation $\rel{M}$ which describes the subset of input elements, where the relation $\rel{H}$ that simulates the head of the machine is forced to spread.

\begin{figure}[ht]
\begin{subfigure}{0.49\textwidth}
\centering
\includegraphics[scale = 1]{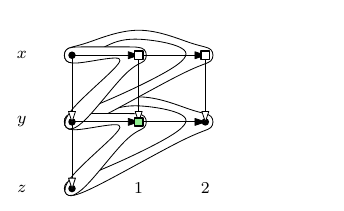}
\caption{Start.}
\label{fig:marked2}
\end{subfigure}
\begin{subfigure}{0.49\textwidth}
\centering
\includegraphics[scale = 1]{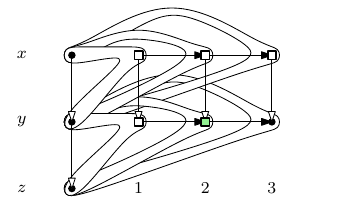}
\caption{Middle.}
\label{fig:marked1}
\end{subfigure}
\begin{subfigure}{0.49\textwidth}
\centering
\includegraphics[scale = 1]{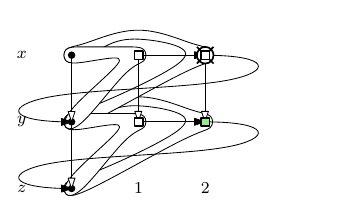}
\caption{End, no head.}
\label{fig:marked3} 
\end{subfigure}
\begin{subfigure}{0.49\textwidth}
\centering
\includegraphics[scale = 1]{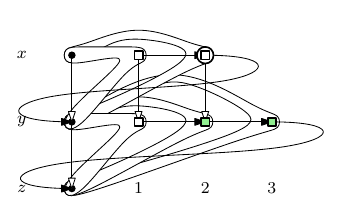}
\caption{End, head.}
\label{fig:marked4}
\end{subfigure}
\caption{All the four cases when $\rel{M}$ (denoted by $\Box$) is forced to spread down. Crossed white circle at \Cref{fig:marked3} stands for a negated $\rel{H}$-atom. White circle at \Cref{fig:marked4} stands for a non-negated $\rel{H}$-atom. The $\rel{M}$-atoms in the head of each implication are coloured in green.}
\label{fig:marked_all1}
\end{figure}

The relation $\rel{M}$ is initiated by the $\rel{I}$-elements:
\begin{equation}\label{eq:initial_to_marked}
\rel{I}(x)\to\rel{M}(x)
\end{equation}
Then $\rel{M}$ must spread along the vertical $\rel{next}$-arcs from top to bottom according to the four rules displayed on Figure~\ref{fig:marked_all1}.
The $\rel{next}$-neighbour must be in $\rel{M}$ only if its $\rel{next}$-predecessor is in $\rel{M}$ and if the structure locally looks like a grid from \Cref{subsection:space_time_diagram}.
Every rule of Figure~\ref{fig:marked_all1} is also given as a logical formula
\begin{align}
&\mu_\mathrm{A}(x,x_1,x_2,y,y_1,y_2,z)\rightarrow\rel{M}(y_1),\quad &&\mu_\mathrm{B}(x,x_1,x_2,x_3,y,y_1,y_2,y_3,z)\rightarrow\rel{M}(y_2),\label{eq:marked1}\\
&\mu_\mathrm{C}(x,x_1,x_2,y,y_1,y_2,z)\rightarrow\rel{M}(y_2),\quad &&\mu_\mathrm{D}(x,x_1,x_2,y,y_1,y_2,y_3,z)\rightarrow\rel{M}(y_2)\wedge\rel{M}(y_3),\label{eq:marked2}
\end{align}
where

\begin{align}
&\mu_\mathrm{A}(x,x_1,x_2,y,y_1,y_2,z) \equiv \rel{succ}(x,x_1)\wedge\rel{succ}(x_1,x_2)\wedge\rel{succ}(y,y_1)\wedge\rel{succ}(y_1,y_2)\nonumber\\
&\wedge\;\rel{next}(x,y)\wedge\rel{next}(y,z)\wedge\rel{next}(x_1,y_1)\wedge\rel{next}(x_2,y_2)\nonumber\\
&\wedge\;\rel{row}(x,x_1,y)\wedge\rel{row}(x,x_2,y)\wedge\rel{row}(y,y_1,z)\wedge\rel{row}(y,y_2,z)\nonumber\\
&\wedge\;\rel{M}(x_1)\wedge\rel{M}(x_2)\label{eq:marked_a}\\
&\nonumber\\
&\mu_\mathrm{B}(x,x_1,x_2,x_3,y,y_1,y_2,y_3,z) \equiv\rel{succ}(x_1,x_2)\wedge\rel{succ}(x_2,x_3)\wedge\rel{succ}(y_1,y_2)\wedge\rel{succ}(y_2,y_3)\nonumber\\
&\wedge\;\rel{next}(x,y)\wedge\rel{next}(y,z)\wedge\rel{next}(x_1,y_1)\wedge\rel{next}(x_2,y_2)\wedge\rel{next}(x_3,y_3)\nonumber\\
&\wedge\;\rel{row}(x,x_1,y)\wedge\rel{row}(x,x_2,y)\wedge\rel{row}(x,x_3,y)\nonumber\\
&\wedge\;\rel{row}(y,y_1,z)\wedge\rel{row}(y,y_2,z)\wedge\rel{row}(y,y_3,z)\nonumber\\
&\wedge\;\rel{M}(x_1)\wedge\rel{M}(x_2)\wedge\rel{M}(x_3)\wedge\rel{M}(y_1)\label{eq:marked_b}\\
&\nonumber\\
&\mu_\mathrm{C}(x,x_1,x_2,y,y_1,y_2,z) \equiv\rel{succ}(x_1,x_2)\wedge\rel{succ}(x_2,y)\wedge\rel{succ}(y_1,y_2)\wedge\rel{succ}(y_2,z)\nonumber\\
&\wedge\;\rel{next}(x,y)\wedge\rel{next}(y,z)\wedge\rel{next}(x_1,y_1)\wedge\rel{next}(x_2,y_2)\nonumber\\
&\wedge\;\rel{row}(x,x_1,y)\wedge\rel{row}(x,x_2,y)\wedge\rel{row}(y,y_1,z)\wedge\rel{row}(y,y_2,z)\nonumber\\
&\wedge\;\neg\rel{H}(x_2)\wedge\rel{M}(x_1)\wedge\rel{M}(x_2)\wedge\rel{M}(y_1)\label{eq:marked_c}\\
&\nonumber\\
&\mu_\mathrm{D}(x,x_1,x_2,y,y_1,y_2,y_3,z) \equiv\nonumber\\
&\equiv\rel{succ}(x_1,x_2)\wedge\rel{succ}(x_2,y)\wedge\rel{succ}(y_1,y_2)\wedge\rel{succ}(y_2,y_3)\wedge\rel{succ}(y_3,z)\nonumber\\
&\wedge\;\rel{next}(x,y)\wedge\rel{next}(y,z)\wedge\rel{next}(x_1,y_1)\wedge\rel{next}(x_2,y_2)\nonumber\\
&\wedge\;\rel{row}(x,x_1,y)\wedge\rel{row}(x,x_2,y)\wedge\;\rel{row}(y,y_1,z)\wedge\rel{row}(y,y_2,z)\wedge\rel{row}(y,y_3,z)\nonumber\\
&\wedge\;\rel{H}(x_2)\wedge\rel{M}(x_1)\wedge\rel{M}(x_2)\wedge\rel{M}(y_1)\label{eq:marked_d}
\end{align}

\subsubsection*{Simulating the transitions of the machine}

Both families of relations $\rel{S}(\cdot)$, for $s\in\Sigma$, and $\rel{Q}(\cdot)$, for $q\in\family{Q}$, partition the elements of the input.
\begin{equation}\label{eq:s_partition}
\neg\left(\bigwedge_{s\in\Sigma}\neg\rel{S}(x)\wedge\rel{M}(x)\right)\wedge\bigwedge_{s,s'\in\Sigma}\neg\bigl(\rel{S}(x)\wedge\rel{S'}(x)\wedge\rel{M}(x)\bigr)
\end{equation}
\begin{equation}\label{eq:q_partition}
\neg\left(\bigwedge_{q\in\family{Q}}\neg\rel{Q}(x)\wedge\rel{M}(x)\right)\wedge\bigwedge_{q,q'\in\family{Q}}\neg\bigl(\rel{Q}(x)\wedge\rel{Q'}(x)\wedge\rel{M}(x)\bigr)
\end{equation}

Moreover, no two elements of the same row can belong to distinct $\rel{Q}$ and $\rel{Q}'$:
\begin{equation}\label{eq:same_row}
\bigwedge_{q,q'\in\family{Q}}\neg\bigl(\rel{row}(a,x,a')\wedge\rel{row}(a,y,a')\wedge\rel{Q}(x)\wedge\rel{Q'}(y)\wedge\rel{M}(x)\wedge\rel{M}(y)\bigr).
\end{equation}

\begin{figure}[ht]
\begin{subfigure}{0.49\textwidth}
\centering
\includegraphics[scale = 1]{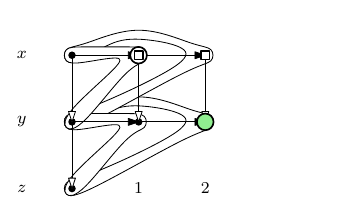}
\caption{Left end.}
\label{fig:head1}
\end{subfigure}
\begin{subfigure}{0.49\textwidth}
\centering
\includegraphics[scale = 1]{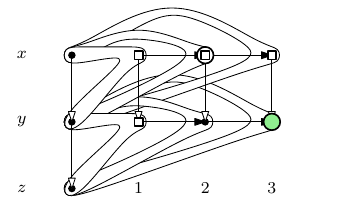}
\caption{Middle. Going right.}
\label{fig:head2}
\end{subfigure}
\begin{subfigure}{0.49\textwidth}
\centering
\includegraphics[scale = 1]{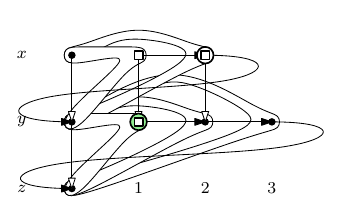}
\caption{Right end.}
\label{fig:head3} 
\end{subfigure}
\begin{subfigure}{0.49\textwidth}
\centering
\includegraphics[scale = 1]{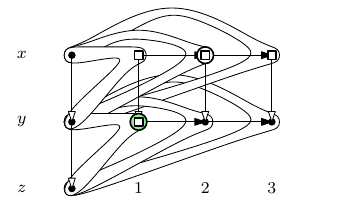}
\caption{Middle. Going left.}
\label{fig:head4}
\end{subfigure}
\caption{All the four cases when $\rel{H}$ (denoted by a circle) is forced to spread. The $\rel{H}$-atoms in the head of each implication are coloured in green.}
\label{fig:head_all}
\end{figure}

After being initiated, the relation $\rel{H}$ spreads down according to the movement of the head of $M$.
As the machine $M$ is back-and-forth, the states of $M$ that are not halting states can be divided in two groups: $\family{Q}^\rightarrow$ and $\family{Q}^\leftarrow$.
If the head of the machine $M$ is in a state from $\family{Q}^\rightarrow$ and if it is not at the right end of the tape, then it keeps moving to the right.
Similarly, if it is in a state from $\family{Q}^\leftarrow$ and if it is not at the left end of the tape, then it keeps moving to the left.
Otherwise, the head is at the left or at the right end of the tape and it has to start moving in the opposite direction.
Notice that, when the head of the machine $M$ is at the first symbol $\triangleright$, it is already determined what it has to write in the cell and move to the right, so only the state can change.
Similarly, when the head is at the blank symbol $\sqcup$, it will move to the left.
Therefore, the transition function $\Delta$ of $M$ can be described by the four functions
\begin{compactenum}[(a)]
  \item $\Delta_\triangleright\colon \family{Q}^\leftarrow \to 2^{\family{Q}^\rightarrow}$, when the head is at the first cell,
  \item $\Delta_\rightarrow\colon \Sigma\times\family{Q}^\rightarrow \to 2^{\Sigma\times\family{Q}^\rightarrow}$, when the head is moving right,
  \item $\Delta_\sqcup\colon \family{Q}^\rightarrow \to 2^{\Sigma\times\family{Q}^\leftarrow}$, when the head is at the right end of the tape,
  \item $\Delta_\leftarrow\colon \Sigma\times\family{Q}^\leftarrow \to 2^{\Sigma\times\family{Q}^\leftarrow}$, when the head is moving left.
\end{compactenum}
Let $\Delta_{\hookleftarrow}\colon \Sigma\times\family{Q}^\rightarrow\to 2^{\Sigma\times\Sigma\times\family{Q}^\leftarrow}$ return, for a pair $(s_0,q_0)\in\Sigma\times\family{Q}^\rightarrow$, the set of all triples $(s_2,s_3,q_3)\in\Sigma\times\Sigma\times\family{Q}^\leftarrow$ such that
\begin{multline}
\exists s_1\in\Sigma, q_1\in\family{Q}^\rightarrow,q_2\in\family{Q}^\leftarrow \\ (s_1,q_1)\in\Delta_\rightarrow(s_0,q_0)\wedge (s_2,q_2)\in\Delta_\sqcup(q_1)\wedge(s_3,q_3)\in\Delta_\leftarrow(s_1,q_2)
\end{multline}
Similarly, let $\Delta_{\hookrightarrow}\colon \Sigma\times\family{Q}^\leftarrow\to 2^{\Sigma\times\family{Q}^\rightarrow}$ return, for a pair $(s_0,q_0)\in\Sigma\times\family{Q}^\leftarrow$, the set of all pairs $(s_2,q_3)\in\Sigma\times\family{Q}^\rightarrow$ such that
\begin{multline}
\exists s_1\in\Sigma,q_1\in\family{Q}^\leftarrow,q_2\in\family{Q}^\rightarrow \\ (s_1,q_1)\in\Delta_\leftarrow(s_0,q_0)\wedge q_2\in\Delta_\triangleright(q_1)\wedge (s_2, q_3)\in\Delta_\rightarrow(s_1,q_2)
\end{multline}
With $\Delta_{\hookleftarrow}$ and $\Delta_{\hookrightarrow}$, the relation $\rel{H}$ does not have to simulate the moments when the head is at the left or at the right end of the tape, it changes its moving direction right before it comes to $\triangleright$ or to $\sqcup$ respectively.
The relation $\rel{H}$ is forced to simulate the movement of the head of the machine only if the underlying element is in $\rel{M}$ and if locally the structure looks like a grid from \Cref{subsection:space_time_diagram}.
Let $\mu_\mathrm{A},\mu_\mathrm{B},\mu_\mathrm{D}$ be as in \cref{eq:marked_a,eq:marked_b,eq:marked_d}.
Then, the execution of $M$ is simulated according to \cref{eq:transition_a,eq:transition_b,eq:transition_c,eq:transition_d}, see \Cref{fig:head_all}.
Each of \cref{eq:transition_a,eq:transition_b,eq:transition_c,eq:transition_d} contains a disjunction of conjunctions in the head of the implication.
It is not allowed by $\MMSNPineq$ but each such clause can be transformed to an allowed form as follows:
\begin{equation*}
A\rightarrow ((B\wedge C)\vee (D\wedge E))\equiv \neg(A\wedge (\neg B\vee \neg C)\wedge(\neg D\vee\neg E))\equiv
\end{equation*}
\begin{equation*}
\equiv \neg(A\wedge \neg B\wedge \neg D)\wedge\neg(A\wedge \neg B\wedge \neg E)\wedge\neg(A\wedge \neg C\wedge \neg D)\wedge\neg(A\wedge \neg C\wedge \neg E)
\end{equation*}

\begin{multline}\label{eq:transition_a}
\bigwedge_{(s,q)\in\Sigma\times\family{Q}^\leftarrow}\biggl(\Bigl(\mu_\mathrm{A}(x,x_1,x_2,y,y_1,y_2,z)\wedge\rel{H}(x_1)\wedge\rel{S}(x_1)\wedge\rel{Q}(x_1)\Bigr)\\
\rightarrow \Bigl(\rel{H}(y_2)\wedge \bigvee_{(s',q')\in\Delta_{\hookrightarrow}(s,q)} \rel{S}'(y_1)\wedge\rel{Q}'(y_2)\Bigr)\biggr)
\end{multline}
\begin{multline}\label{eq:transition_b}
\bigwedge_{(s,q)\in\Sigma\times\family{Q}^\rightarrow}\biggl(\Bigl(\mu_\mathrm{B}(x,x_1,x_2,x_3,y,y_1,y_2,y_3,z)\wedge\rel{H}(x_2)\wedge\rel{S}(x_2)\wedge\rel{Q}(x_2)\Bigr)\\
\rightarrow \Bigl(\rel{H}(y_3)\wedge \bigvee_{(s',q')\in\Delta_\rightarrow(s,q)} \rel{S}'(y_2)\wedge\rel{Q}'(y_3)\Bigr)\biggr)
\end{multline}
\begin{multline}\label{eq:transition_c}
\bigwedge_{(s,q)\in\Sigma\times\family{Q}^\rightarrow}\biggl(\Bigl(\mu_\mathrm{D}(x,x_1,x_2,y,y_1,y_2,y_3,z)\wedge\rel{H}(x_2)\wedge\rel{S}(x_2)\wedge\rel{Q}(x_2)\Bigr)\\
\rightarrow \Bigl(\rel{H}(y_1)\wedge\bigvee_{(s'_\sqcup,s',q')\in\Delta_{\hookleftarrow}(s,q)} \rel{S}'(y_2)\wedge\rel{S}'_\sqcup(y_3)\wedge\rel{Q}'(y_1)\Bigr)\biggr)
\end{multline}
\begin{multline}\label{eq:transition_d}
\bigwedge_{(s,q)\in\Sigma\times\family{Q}^\leftarrow}\biggl(\Bigl(\mu_\mathrm{B}(x,x_1,x_2,x_3,y,y_1,y_2,y_3,z)\wedge\rel{H}(x_2)\wedge\rel{S}(x_2)\wedge\rel{Q}(x_2)\Bigr)\\
\rightarrow \Bigl(\rel{H}(y_1)\wedge\bigvee_{(s',q')\in\Delta_\leftarrow(s,q)} \rel{S}'(y_2)\wedge\rel{Q}'(y_1)\Bigr)\biggr)
\end{multline}

If an element $x$ does not belong to $\rel{H}$, then its $\rel{next}$-successor must have the same symbol relation as $x$ has:
\begin{equation}\label{eq:symbol_propagate}
\bigwedge_{s\in\Sigma}\bigl(\rel{M}(x)\wedge\neg\rel{H}(x)\wedge\rel{S}(x)\wedge\rel{next}(x,x')\to\rel{S}(x')\bigr)
\end{equation}

The head cannot visit two distinct cells at the same moment of  time, so we need the following rule:
\begin{equation}\label{eq:head_unique}
\rel{row}(a,x,a')\wedge\rel{row}(a,y,a')\wedge\rel{M}(x)\wedge\rel{M}(y)\wedge\rel{H}(x)\wedge\rel{H}(y)\to x=y
\end{equation}

Let $\rel{Q}_r$ be a relation associated with a rejecting state $q_r$ in $\family{Q}$. Then, for every such $\rel{Q}_r$, we add the following rule that forbids a head to be in this state during the execution:
\begin{equation}\label{eq:rejecting_state}
\neg\bigl(\rel{M}(x)\wedge\rel{H}(x)\wedge\rel{Q}_r(x)\bigr)
\end{equation}

\subsubsection*{First-order restrictions}
The expressiveness of $\MMSNPineq$ permits $\Phi$ to require the following properties that reject \emph{over-complete} input structures. Notice that the following negated conjuncts have no $\sigma$-atoms (existentially quantified). Therefore, we will call them \emph{first-order restrictions}.
\begin{compactenum}[(a)]
\item There is at most one $\rel{start}$ element: $\rel{start}(x)\wedge\rel{start}(y)\rightarrow x=y$.
\item The in- and out- degrees of $\rel{succ}$ and $\rel{next}$ are at most 1:
\begin{equation*} 
(\rel{succ}(x,y)\wedge\rel{succ} (x,y')\rightarrow y=y')\wedge(\rel{succ}(x,y)\wedge\rel{succ} (x',y)\rightarrow x=x')
\end{equation*} 
\begin{equation*} 
\wedge(\rel{next}(x,y)\wedge\rel{next}(x,y')\rightarrow y=y')\wedge (\rel{next} (x,y)\wedge\rel{next}(x',y)\rightarrow x=x').
\end{equation*}
\item An element $y$ cannot be in different triples $\rel{row}(x,y,z)$ and $\rel{row}(x',y,z')$:
\begin{equation*} 
(\rel{row}(x,y,z)\wedge\rel{row} (x',y,z')\rightarrow x=x')
\wedge(\rel{row}(x,y,z)\wedge\rel{row}(x',y,z')\rightarrow z=z').
\end{equation*}
\item Loops $\rel{succ}(x,x),\rel{next}(x,x)$ are forbidden as well as $\rel{row}(x,y,x)$.
\item An element $x$ is contained in at most one symbol relation $\rel{s}(x)$: 
\begin{equation*}
\bigwedge_{\text{distinct }s,s'\in\Sigma}\neg\bigl(\rel{s}(x)\wedge\rel{s}' (x)\bigr).
\end{equation*}
\end{compactenum}

\subsection{Reduction from $M$ to $\Phi$}
Let $M$ be a precise Turing machine and let $\Phi$ be a $\MMSNPineq$ sentence constructed from $M$ according to Subsection~\ref{subsection:construction_of_phi}. Let $\func{r_{easy}}$ be a mapping from the set of finite strings $\Sigma^*$ to the set of finite $\tau$-structures defined in Subsection~\ref{subsection:space_time_diagram}.
\begin{lem}\label{lemma:np_reduces_to_mmsnpineq}
The mapping $\func{r_{easy}}$ describes a polynomial-time reduction from $M$ to $\Phi$.
\end{lem}
\proof
Let $\tuple{s}\in\Sigma^\ast$ be an input string of $M$ and let $\structure{A}=\func{r_{easy}}(\tuple{s})$.

\begin{figure}
\centering
\includegraphics[width=0.5\textwidth]{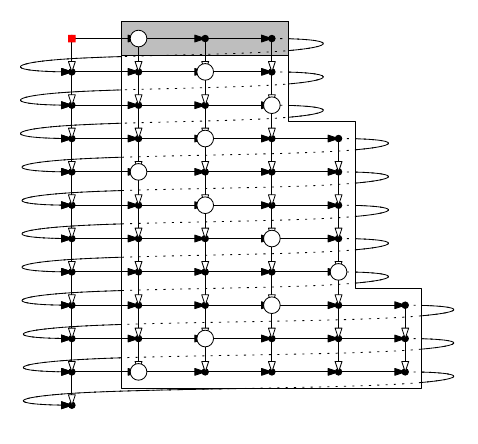}
\caption{The unique choices for the relations $\rel{I}$ (grey), $\rel{M}$ (grey and white), and $\rel{H}$(white circles) for the structure from Example~\ref{ex:construction}.}
\label{fig:m_to_phi_example}
\end{figure}

By the construction of $\Phi$, there are unique interpretations of $\rel{M},\rel{H},\rel{I}$ that satisfy the rules of $\Phi$ and the relation $\rel{H}$ simulates exactly $\func{g}(n)$ round trips of the head of $M$, see Figure~\ref{fig:m_to_phi_example} for example.
By the construction of $\Phi$, there exists a one-to-one correspondence between transitions of the head of the machine and the set of rules of $\Phi$ that describe these transitions. 
So, if the head of the machine is in a rejecting state at the end, then, for every possible choice of existentially quantified relations, the rule (\ref{eq:rejecting_state}) will be violated. 
Otherwise, if there exists a sequence of transitions that finishes in an accepting state, then the construction of $\Phi$ allows to repeat exactly the same sequence, so the rule (\ref{eq:rejecting_state}) will not be violated.
\qed

\subsection{Reduction from $\Phi$ to $M$}
It remains to describe the mapping $\func{r_{hard}}$ that takes as input a finite $\tau$-structure $\structure{A}$ and outputs in time polynomial in $|A|$ a string $\tuple{s}\in\Sigma^*$ such that the precise machine $M$ accepts $\tuple{s}$ if and only if $\structure{A}\models\Phi$.

\subsubsection*{Rejecting over-complete input}

One can check whether the input structure $\structure{A}$ violates one of the first-order restrictions in polynomial time in the size of $\structure{A}$. If so, then the mapping $\func{r_{hard}}$ assigns to $\structure{A}$ the fixed NO instance $\tuple{s}_{\mathrm{N}}$.

\subsubsection*{Unique choices for the relations $\rel{M}$ and $\rel{H}$}

Now we may assume that $\structure{A}$ satisfies the first-order restrictions.
We describe the minimal possible sets $M_{\min}$ and $H_{\min}$ consisting of elements of $\structure{A}$ that ought to be in the relations $\rel{M}$ and $\rel{H}$ in order to satisfy $\Phi$.
We will show in Lemma~\ref{claim:marked_minimal} that without loss of generality we can consider exclusively the elements from these sets and  assume that the relations $\rel{M}$ and $\rel{H}$ are interpreted as $M_{\min}$ and $H_{\min}$ respectively.

Denote by $a_0$ the element of $\structure{A}$ such that $\structure{A}\models\rel{start}(a_0)$.
If such an element does not exist, then no element is forced to be in $\rel{M}$, so $\structure{A}$ is immediately accepted by $\Phi$ and $\func{r_{hard}}$ can assign to $\structure{A}$ the string $\tuple{s}_\mathrm{Y}$.
Otherwise, the sets $M_{\min}$ and $H_{\min}$ are defined by induction.

Firstly, we add to $M_{\min}$ the right $\rel{succ}$-neighbour of $a_0$ according to \cref{eq:initial1}. 
Then we add to $M_{\min}$ the neighbour of the neighbour of $a_0$, and so on, by using the rule from \cref{eq:initial2}.
Eventually, we add all the elements that are forced to be in $\rel{I}$.
By \cref{eq:initial_to_marked}, every element in $\rel{I}$ must also be in $\rel{M}$.
Then, we add to $M_{\min}$ every element that is forced to be in $\rel{M}$ according to \cref{eq:marked1,eq:marked2}, and add to $H_{\min}$ every element that is forced to be in $\rel{H}$ according to \cref{eq:transition_a,eq:transition_b,eq:transition_c,eq:transition_d}.
Repeat this until we do not have to add any new element to $M_{\min}$ and $H_{\min}$.

As $\structure{A}$ satisfies the first-order restrictions, the in- and out-degrees of $\rel{succ}$- and $\rel{next}$-arcs of $\structure{A}$ cannot be greater than 1.
So, we can assign a pair of natural numbers to every element of $M_{\min}$.
Assign $(1,0)$ to the right $\rel{succ}$-neighbour of $a_0$.
If $(i,j)$ is assigned to $x$ and if $\structure{A}\models\rel{succ}(x,y)$, then we assign $(i+1,j)$ to $y$.
If $(i,j)$ is assigned to $x$ and if $\structure{A}\models\rel{next}(x,y)$, then we assign $(i,j+1)$ to $y$.
It follows from the definition of $M_{\min}$ and from the assumption that $\structure{A}$ satisfies the first-order restrictions, that this assignment is well-defined.

Let $S$ be the set of $\rel{next}$-successors of $a_0$.
That is, $a_0$ belongs to $S$ and, if $\structure{A}\models\rel{next}(x,y)$ and $x$ is in $S$, then $y$ belongs to $S$.
Let $\structure{A}_{\min}$ be the substructure of $\structure{A}$ induced on $A_{\min}:=S\cup M_{\min}$.

\begin{lem}\label{claim:marked_minimal} 
Let $\structure{A}$ satisfy the first-order restrictions of $\Phi$.
Then, $\Phi$ holds in $\structure{A}$ if and only if it holds in $\structure{A}_{\min}$.
\end{lem}
\proof
It is sufficient to show only that $\structure{A}_{\min}\models\Phi$ implies $\structure{A}\models\Phi$.
The other direction is clear because SNP problems are closed under taking substructures.

\begin{figure}
\centering
\includegraphics[width=0.45\textwidth]{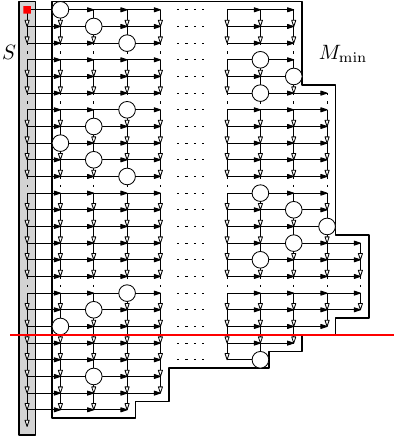}
\caption{A $\sigma$-expansion of the substructure of $\structure{A}$ induced on $M_{\min}$ that satisfies the first-order part of $\Phi$.
Grey zone highlight the set $S$.
The red square is $\rel{start}$, horizontal arcs are $\rel{succ}$, vertical are $\rel{next}$.
Circles are the elements of $\rel{H}$.
The red line highlights the moment of time when $M$ halts.}
\label{fig:cm_ch}
\end{figure}

Let $\structure{A}_{\min}^\sigma$ be a $\sigma$-expansion of $\structure{A}_{\min}$ that satisfies the first-order part of $\Phi$.
An example of such a structure $\structure{A}_{\min}^\sigma$ is given on Figure~\ref{fig:cm_ch}.
By the construction of $M_{\min}$ and $H_{\min}$, we know that $M_{\min}\subseteq\rel{M}^{\structure{A}_{\min}^\sigma}$ and  that $H_{\min}\subseteq \rel{H}^{\structure{A}_{\min}^\sigma}$.
We can assume without loss of generality that $\rel{M}^{\structure{A}_{\min}^\sigma}=M_{\min}$ because we can just remove every element of $(A_{\min}\setminus M_{\min})$ from $\rel{M}^{\structure{A}_{\min}^\sigma}$ without violating a rule of $\Phi$.

Let $\structure{A}^\sigma$ be any $\sigma$-expansion of $\structure{A}$ that contains $\structure{A}^\sigma_{\min}$ as a substructure and such that $\rel{M}^{\structure{A}^\sigma}= \rel{M}^{\structure{A}_{\min}^\sigma} = M_{\min}$ and $\rel{H}^{\structure{A}^\sigma} = \rel{H}^{\structure{A}_{\min}^\sigma}$.
By the definition of $M_{\min}$, $\structure{A}^{\sigma}$ satisfies all the rules from \cref{eq:initial1,eq:initial2,eq:initial_to_marked,eq:marked1,eq:marked2}.
In all other conjuncts except for the variables $a$ and $a'$ of (\ref{eq:same_row}) and (\ref{eq:head_unique}), all variables are assumed to be in $\rel{M}$, so such conjuncts will not be violated in $\structure{A}^\sigma$.
Suppose that one of the negated conjuncts from (\ref{eq:same_row}) or (\ref{eq:head_unique}) is violated.
Let $a,a',x$, and $y$ witness it.
As $x$ and $y$ are in $\rel{M}$, they must be in $M_{\min}$.
So, at least one of $a$ or $a'$ is not in $A_{\min}$.
But, by the definition of $M_{\min}$, if $x$ and $y$ are in $M_{\min}$, then both $a$ and $a'$ are in $S$.
So, all of them belong to $A_{\min}$, this contradicts $\structure{A}_{\min}\models\Phi$.
\qed

\subsubsection*{The reduction $\func{r_{hard}}$}
We are ready to construct the mapping $\func{r_{hard}}$ that witnesses a polynomial-time reduction from $\Phi$ to $M$.

If a $\tau$-structure $\structure{A}$ does not satisfy the first-order restrictions, then $\func{r_{hard}}(\structure{A})=\tuple{s}_{\mathrm{N}}$, hence, it is also rejected by $M$.
Otherwise, by Lemma~\ref{claim:marked_minimal}, we can assume that $\structure{A} = \structure{A}_{\min}$.
Suppose that the first row of $M_{\min}$ consists of the elements $(1,0),\ldots,(n,0)$.
The construction of the sets $M_{\min}$ and $H_{\min}$ is deterministic and requires polynomial time in the size of $\structure{A}$.

\begin{figure}[h]
\begin{subfigure}{0.45\textwidth}
\includegraphics[width=\textwidth]{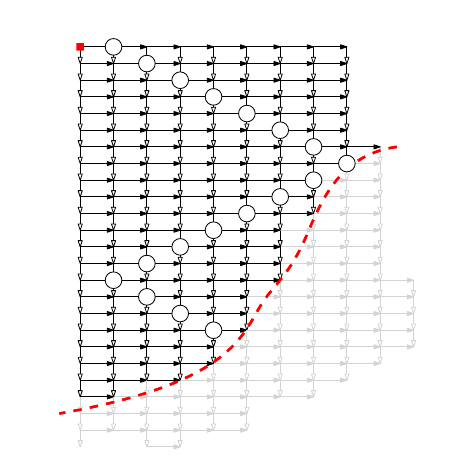}
\caption{An incomplete structure.}
\label{fig:difficult2}
\end{subfigure}
\begin{subfigure}{0.45\textwidth}
\includegraphics[width=\textwidth]{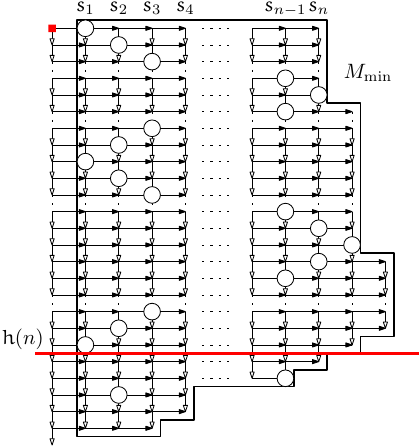}
\caption{A complete structure.}
\label{fig:difficult3}
\end{subfigure}
\label{fig:difficult}
\end{figure}

Recall from \cref{subsection:space_time_diagram} that, in order to simulate every execution of $M$, the set $M_{\min}$ must have at least $\func{h}(n) = \bigl(\func{g}(n)\bigr)^2 + (2n-3)\func{g}(n) + 1$ rows.
If the set $H_{\min}$ does not contain the element $\bigl(1,\func{h}(n)\bigr)$, then we know that $\structure{A}$ cannot simulate the whole execution of $M$, so the head will never be in a rejecting state, \ie, the rule (\ref{eq:rejecting_state}) will not be violated.
See Figure~\ref{fig:difficult2} for an illustration.
Other rules can be satisfied for the $\sigma$-expansion $\structure{A}^\sigma$, where $\rel{M}^{\structure{A}^\sigma} = M_{\min}$, $\rel{H}^{\structure{A}^\sigma}=H_{\min}$, and the relations $\rel{S}$ (for every $s\in\Sigma$) and $\rel{Q}$ (for every $q\in\family{Q}$) are chosen in such a way that no rule among \cref{eq:initiating_q,eq:initiating_s,eq:s_partition,eq:q_partition,eq:same_row,eq:transition_a,eq:transition_b,eq:transition_c,eq:transition_d,eq:symbol_propagate} is violated.
In this case, $\func{r_{hard}}(\structure{A}) = \tuple{s}_{\mathrm{Y}}$.

If the set $H_{\min}$ contains the element $\bigl(1,\func{h}(n)\bigr)$, then we know that $\structure{A}$ allows to simulate the whole execution of $M$ on the input of size $n$.
Define $\func{r_{hard}}(\structure{A})$ to be a string $\tuple{s}=s_1\ldots s_n$ such that, for every $i\in[n]$, the $i$th element $s_i$ is equal to some $s$ in $\Sigma$ if and only if the element $(i,0)$ is in $\rel{s}^\structure{A}$.
See Figure~\ref{fig:difficult3} for an illustration.
As $\structure{A}$ satisfies the first-order restrictions, no element of $\structure{A}$ belongs to two distinct ``symbol'' relations $\rel{s}_1^\structure{A}$ and $\rel{s}_2^\structure{A}$.
As each element of the first row is in $M_{\min}$, it must belong to at least one ``symbol'' relation $\rel{s}^\structure{A}$.
So, the mapping $\func{r_{hard}}$ is well-defined.
The following lemma directly follows from the definition of $\func{r_{hard}}$.
\begin{lem}\label{lemma:mmsnpineq_reduces_to_np} 
For every $\tau$-structure $\structure{A}$, the machine $M$ accepts $\func{r_{hard}}(\structure{A})$ if and only if $\structure{A}\models\Phi$.
\end{lem}

Lemma~\ref{lemma:mmsnpineq_reduces_to_np} together with Lemma~\ref{lemma:np_reduces_to_mmsnpineq} imply Theorem~\ref{th:np_subset_mmsnpineq}.

\section{Conclusion}

In this paper, we have explored some fragments that sit above Feder and Vardi's logic MMSNP and below the three logics obtained by relaxing one of the three syntactic restrictions: namely, (non monotone) monadic $\SNP$, $\MMSNP$ with $\not=$ and (non monadic) monotone $\SNP$.
These three logics cannot exhibit a dichotomy unless P is equal to NP by Ladner's theorem.
In order to extend a dichotomy to some superclasses of $\MMSNP$, we have highlighted three more general fragments thereof: namely, $\MPART$, $\GMMSNPineq$ and $\GMSNP$.
Apart from $\GMMSNPineq$ which we embed up to polynomial-time equivalence in $\MMSNP$ to prove dichotomy, the question remains open for its two siblings.

A weaker yet interesting result would consist in relating directly the dichotomy status among two classes for which the dichotomy question is open.
For example, is it the case that for every problem in $\MPART$ there exists a polynomial-time equivalent problem in $\GMSNP$?

Another puzzling question concerns the the class $\MPART$ that contains the variation of Matrix Partitions (MP) defined in~\cite{barsukovkante}.
In particular is it the case that one can find for each sentence of $\MPART$ a polynomial-time equivalent problem in MP?
The analogous question for CSP and MMSNP was settled by Feder and Vardi using Erd\H{o}s' lemma which allows to construct in randomised polynomial-time a structure of high girth and fixed chromatic number. 
The technique breaks down in the context of $\MPART$.
Indeed, the main difficulty when working in the context of MP and $\MPART$ is that it is a genuine challenge to combine structures as is common when working in the monotone world of CSP and MMSNP. 
In the latter, we are free to take the disjoint union of accepted instances to produce a larger accepted instance or to combine two accepted instances sharing some common feature (\eg~free amalgamation). 
In the former, absence of an edge between vertices does not usually mean that we may freely map these vertices.
While it is possibly the lack of freedom of this feature that means that the problems could not show NP-intermediate complexity, it poses a real challenge when working with these non monotone problems.

If we could somehow show that each problem in  $\GMSNP$ enjoys a polynomial-time equivalent problem in MMSNP, then we would obtain a dichotomy for $\GMSNP$.
The challenge in this endeavour is that some input structures in the target problem of MMSNP are not in the image of the reduction from $\GMSNP$. 
This is precisely the problem that was solved by Feder and Vardi using Erd\H{o}s' Lemma in the context of MMSNP and CSP.  

The above question seems hard and a perhaps weaker question that could serve as a first step would consist in showing that a problem in $\GMSNP$ over a signature with symbols of arity at most $r$ enjoys a polynomial-time equivalent problem in $\GMSNP$ over a signature with symbols of arity at most $r-1$. 
In particular, the case $r=3$ would be an interesting first step.

\section*{Acknowledgment}
  \noindent The authors sincerely appreciate the anonymous referees whose valuable comments helped to improve the presentation and writing of this article.

\bibliographystyle{alphaurl}
\bibliography{lmcs2023}

\end{document}

%% file: mymacros.tex
\DeclareMathAlphabet\EuRoman{U}{eur}{m}{n}
\SetMathAlphabet\EuRoman{bold}{U}{eur}{b}{n}
\newcommand{\eurm}[1]{\EuRoman{#1}} %euler roman

\DeclareMathAlphabet\EuScript{U}{eus}{m}{n}
\SetMathAlphabet\EuScript{bold}{U}{eus}{b}{n}
\newcommand{\euca}[1]{\mathcal{#1}} %euler calligraphic

\newcommand{\eg}{\emph{e.g.}\xspace}  % e.g.
\newcommand{\ie}{i.e.\xspace}  % id est
\newcommand*\guardedineq{\includegraphics{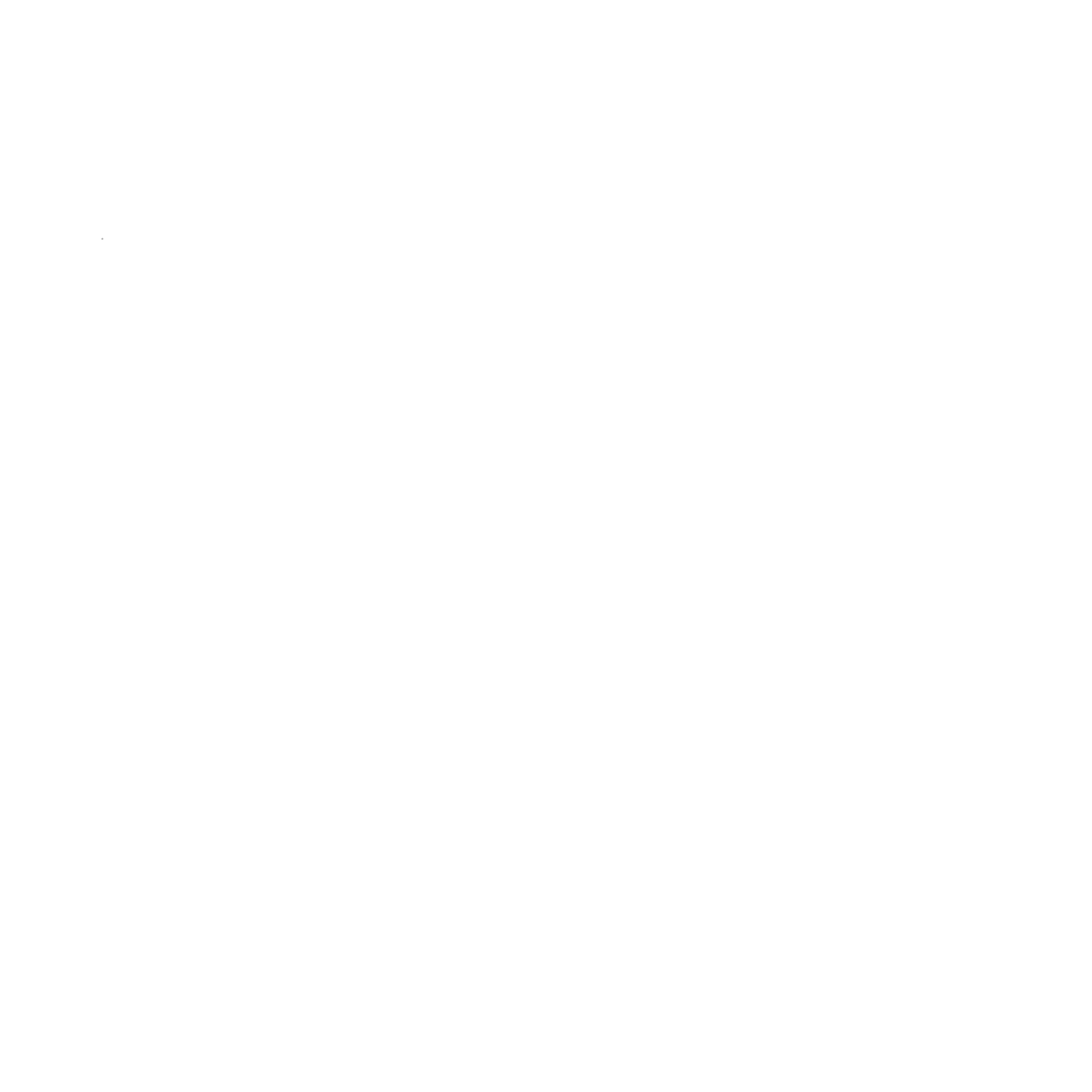}}
\newcommand{\structure}[1]{\mathfrak{#1}} %structures, graphs
\newcommand{\tuple}[1]{\mathbf{#1}} %tuple
\newcommand{\rel}[1]{\eurm{#1}} %relation
\newcommand{\family}[1]{\euca{#1}} %class, family of objects

\newcommand{\func}[1]{\eurm{#1}}  %function

\renewcommand{\P}[0]{\textrm{P}}
\newcommand{\NP}[0]{\textrm{NP}}
\newcommand{\NPc}[0]{\textrm{NP-complete}}

\newcommand{\Sat}[0]{\ensuremath{\mathrm{fm}}}

\newcommand{\SNP}[0]{\textrm{SNP}}

\newcommand{\MMSNPineq}[0]{\ensuremath{\textrm{MMSNP}_{\not=}}}
\newcommand{\MMSNP}[0]{\textrm{MMSNP}}
\newcommand{\MMSNPtwo}[0]{\ensuremath{\textrm{MMSNP}_2}}
\newcommand{\GMSNP}[0]{\textrm{GMSNP}}
\newcommand{\GMMSNPineq}[0]{\textrm{MMSNP}_{\guardedineq}}

\newcommand{\MPART}[0]{\ensuremath{\textrm{MPART}}}
\newcommand{\GMPARTineq}[0]{\textrm{MPART}_{\guardedineq}}

\newcommand{\card}[1]{\vert #1 \vert}  % cardinal, size of a set
\renewcommand{\setminus}{\smallsetminus}

\graphicspath{{figures/} }

%% file: lmcs2023.bbl
\begin{thebibliography}{BtCLW14}

\bibitem[Bar22]{barsukovphd}
Alexey Barsukov.
\newblock {\em On dichotomy above Feder and Vardi's logic}.
\newblock PhD thesis, Université Clermont Auvergne, 2022.

\bibitem[BH90]{bangjensen_hell1990}
J{\o}rgen Bang{-}Jensen and Pavol Hell.
\newblock The effect of two cycles on the complexity of colourings by directed graphs.
\newblock {\em Discret. Appl. Math.}, 26(1):1--23, 1990.
\newblock \href {https://doi.org/10.1016/0166-218X(90)90017-7} {\path{doi:10.1016/0166-218X(90)90017-7}}.

\bibitem[BK24]{barsukovkante}
Alexey Barsukov and Mamadou~Moustapha Kant{\'{e}}.
\newblock Generalisations of matrix partitions: Complexity and obstructions.
\newblock {\em Theor. Comput. Sci.}, 1007:114652, 2024.
\newblock URL: \url{https://doi.org/10.1016/j.tcs.2024.114652}, \href {https://doi.org/10.1016/J.TCS.2024.114652} {\path{doi:10.1016/J.TCS.2024.114652}}.

\bibitem[BKN09]{barto_kozik_niven2009}
Libor Barto, Marcin Kozik, and Todd Niven.
\newblock The {CSP} dichotomy holds for digraphs with no sources and no sinks {(A} positive answer to a conjecture of {Bang-Jensen} and {Hell}).
\newblock {\em {SIAM} J. Comput.}, 38(5):1782--1802, 2009.
\newblock \href {https://doi.org/10.1137/070708093} {\path{doi:10.1137/070708093}}.

\bibitem[BKS20]{bodirsky_asnp}
Manuel Bodirsky, Simon Kn{\"{a}}uer, and Florian Starke.
\newblock {ASNP:} {A} tame fragment of existential second-order logic.
\newblock In Marcella Anselmo, Gianluca~Della Vedova, Florin Manea, and Arno Pauly, editors, {\em CiE 2020, Proceedings}, volume 12098 of {\em Lecture Notes in Computer Science}, pages 149--162. Springer, 2020.
\newblock URL: \url{https://doi.org/10.1007/978-3-030-51466-2\_13}, \href {https://doi.org/10.1007/978-3-030-51466-2_13} {\path{doi:10.1007/978-3-030-51466-2_13}}.

\bibitem[Bod21]{bodirsky_book}
Manuel Bodirsky.
\newblock {\em Complexity of Infinite-Domain Constraint Satisfaction}.
\newblock Lecture Notes in Logic. Cambridge University Press, 2021.
\newblock Postprint available online: \url{https://wwwpub.zih.tu-dresden.de/~bodirsky/Book.pdf}.
\newblock \href {https://doi.org/10.1017/9781107337534} {\path{doi:10.1017/9781107337534}}.

\bibitem[BPP21]{bodirsky_pinsker_conjecture}
Manuel Bodirsky, Michael Pinsker, and Andr{\'{a}}s Pongr{\'{a}}cz.
\newblock Projective clone homomorphisms.
\newblock {\em J. Symb. Log.}, 86(1):148--161, 2021.
\newblock URL: \url{https://doi.org/10.1017/jsl.2019.23}, \href {https://doi.org/10.1017/JSL.2019.23} {\path{doi:10.1017/JSL.2019.23}}.

\bibitem[BtCLW14]{bienvenu2014}
Meghyn Bienvenu, Balder ten Cate, Carsten Lutz, and Frank Wolter.
\newblock Ontology-based data access: {A} study through disjunctive datalog, {CSP}, and {MMSNP}.
\newblock {\em {ACM} Trans. Database Syst.}, 39(4):33:1--33:44, 2014.
\newblock \href {https://doi.org/10.1145/2661643} {\path{doi:10.1145/2661643}}.

\bibitem[Bul17]{bulatov2017}
Andrei~A. Bulatov.
\newblock A dichotomy theorem for nonuniform {CSPs}.
\newblock In Chris Umans, editor, {\em 58th {IEEE} Annual Symposium on Foundations of Computer Science, {FOCS}}, pages 319--330. {IEEE} Computer Society, 2017.
\newblock \href {https://doi.org/10.1109/FOCS.2017.37} {\path{doi:10.1109/FOCS.2017.37}}.

\bibitem[Fag74]{fagin1974}
Ronald Fagin.
\newblock Generalized first-order spectra and polynomial-time recognizable sets.
\newblock In {\em Complexity of computation ({P}roc. {SIAM}-{AMS} {S}ympos., {N}ew {Y}ork, 1973)}, volume Vol. VII of {\em SIAM-AMS Proc.}, pages 43--73. Amer. Math. Soc., Providence, RI, 1974.

\bibitem[FV98]{federvardi1998}
Tom{\'{a}}s Feder and Moshe~Y. Vardi.
\newblock The computational structure of monotone monadic {SNP} and constraint satisfaction: {A} study through datalog and group theory.
\newblock {\em {SIAM} J. Comput.}, 28(1):57--104, 1998.
\newblock \href {https://doi.org/10.1137/S0097539794266766} {\path{doi:10.1137/S0097539794266766}}.

\bibitem[FV03]{federV03}
Tom{\'{a}}s Feder and Moshe~Y. Vardi.
\newblock Homomorphism closed vs. existential positive.
\newblock In {\em 18th {IEEE} Symposium on Logic in Computer Science {(LICS} 2003)}, pages 311--320. {IEEE} Computer Society, 2003.
\newblock \href {https://doi.org/10.1109/LICS.2003.1210071} {\path{doi:10.1109/LICS.2003.1210071}}.

\bibitem[Hel14]{hell2014}
Pavol Hell.
\newblock Graph partitions with prescribed patterns.
\newblock {\em Eur. J. Comb.}, 35:335--353, 2014.
\newblock \href {https://doi.org/10.1016/j.ejc.2013.06.043} {\path{doi:10.1016/j.ejc.2013.06.043}}.

\bibitem[HN90]{hellnesetril1990}
Pavol Hell and Jaroslav Ne{\v s}et{\v r}il.
\newblock On the complexity of \emph{H}-coloring.
\newblock {\em J. Comb. Theory, Ser. {B}}, 48(1):92--110, 1990.
\newblock \href {https://doi.org/10.1016/0095-8956(90)90132-J} {\path{doi:10.1016/0095-8956(90)90132-J}}.

\bibitem[Lad75]{ladner1975}
Richard~E. Ladner.
\newblock On the structure of polynomial time reducibility.
\newblock {\em J. {ACM}}, 22(1):155--171, 1975.
\newblock \href {https://doi.org/10.1145/321864.321877} {\path{doi:10.1145/321864.321877}}.

\bibitem[Mad09]{madelaine2009}
Florent~R. Madelaine.
\newblock Universal structures and the logic of forbidden patterns.
\newblock {\em Log. Methods Comput. Sci.}, 5(2), 2009.
\newblock \href {https://doi.org/10.2168/LMCS-5(2:13)2009} {\path{doi:10.2168/LMCS-5(2:13)2009}}.

\bibitem[Mat93]{matiyasevich}
Y.~V. Matiyasevich.
\newblock {\em Hilbert’s Tenth Problem}.
\newblock The MIT Press, 1993.

\bibitem[Pap94]{papadimitriou_book}
Christos~H. Papadimitriou.
\newblock {\em Computational complexity}.
\newblock Addison-Wesley, 1994.

\bibitem[PF79]{pippenger1979}
Nicholas Pippenger and Michael~J. Fischer.
\newblock Relations among complexity measures.
\newblock {\em J. {ACM}}, 26(2):361--381, 1979.
\newblock \href {https://doi.org/10.1145/322123.322138} {\path{doi:10.1145/322123.322138}}.

\bibitem[Pin22]{pinsker_survey}
Michael Pinsker.
\newblock Current challenges in infinite-domain constraint satisfaction: Dilemmas of the infinite sheep.
\newblock In {\em 52nd {IEEE} International Symposium on Multiple-Valued Logic, {ISMVL} 2022, Dallas, TX, USA, May 18-20, 2022}, pages 80--87. {IEEE}, 2022.
\newblock \href {https://doi.org/10.1109/ISMVL52857.2022.00019} {\path{doi:10.1109/ISMVL52857.2022.00019}}.

\bibitem[Sch78]{schaefer1978}
Thomas~J. Schaefer.
\newblock The complexity of satisfiability problems.
\newblock In Richard~J. Lipton, Walter~A. Burkhard, Walter~J. Savitch, Emily~P. Friedman, and Alfred~V. Aho, editors, {\em Proceedings of the 10th Annual {ACM} Symposium on Theory of Computing, May 1-3, 1978, San Diego, California, {USA}}, pages 216--226. {ACM}, 1978.
\newblock \href {https://doi.org/10.1145/800133.804350} {\path{doi:10.1145/800133.804350}}.

\bibitem[Zhu20]{zhuk2020}
Dmitriy Zhuk.
\newblock A proof of the {CSP} dichotomy conjecture.
\newblock {\em J. {ACM}}, 67(5):30:1--30:78, 2020.
\newblock \href {https://doi.org/10.1145/3402029} {\path{doi:10.1145/3402029}}.

\end{thebibliography}
